\documentclass[entropy,article,accept,moreauthors]{mdpi} 

\firstpage{1} 
\pubvolume{xx}
\issuenum{1}
\articlenumber{5}
\pubyear{2020}
\copyrightyear{2020}
\history{Received: date; Accepted: date; Published: date}
\usepackage{textcomp}
\usepackage{gensymb}
\usepackage{amssymb}
\usepackage{amsthm}
\usepackage{longtable}

\Title{Indistinguishability and Negative Probabilities}

\Author{J. Acacio de Barros
 $^{1,}$*\orcidA{} and Federico Holik $^{2}$\orcidB{}}

\AuthorNames{J. Acacio de Barros and Federico Holik}

\address{%
$^{1}$ \quad School of Humanities and Liberal Studies, San Francisco State University, San Francisco, CA 94132, USA\\
$^{2}$ \quad Instituto de Física La Plata, UNLP, CONICET, Facultad de Ciencias Exactas, 1900 La Plata, Argentina; olentiev2@gmail.com}

\corres{Correspondence: barros@sfsu.edu; Tel.: +1-415-405-2674}

\abstract{In this paper, we examined the connection between quantum systems' indistinguishability and signed (or negative) probabilities. We do so by first introducing a measure-theoretic definition of signed probabilities inspired by research in quantum contextuality. We then argue that ontological indistinguishability leads to the no-signaling condition and negative probabilities.}


\begin{document}

\section{Introduction\label{sec:introduction}}
The assignment of truth values to propositions asserting that a system's property has a definite value is problematic in quantum mechanics. Take the case of propositions about momentum and position for a quantum system. Heisenberg's uncertainty principle asserts that we cannot know the values of position and momentum simultaneously, at least not as precisely as one wants. This~constraint brings the issue of whether systems have well-defined but unknowable values of position and momentum, or whether these are undefined. If the former, the probabilistic uncertainties appearing in quantum theory would have an epistemic character, being quantum properties the best description of what we can say about the system. If the latter, then what properties does the system have? For instance, when we measure a particle's momentum  and find the value $p$, does it mean the particle has momentum $p$? (We stress that the word ``particle'' is used here for clarity and that we are not espousing a particle ontology. Our considerations apply to any quantum entities, such as particles or fields, provided they are indistinguishable) 
Moreover, is this value of momentum something that existed before the measurement? If not, then do measurements create properties? Do the experimenter, who chooses what to measure, set what properties a particle has? These questions become more problematic if we consider the Kochen-Specker theorem. 

In their seminal paper, Kochen and Specker (KS) studied hidden-variable theories compatible with the quantum formalism and satisfying certain physically-motivated conditions. They proved that the values that these hidden variable theories assign to propositions about quantum systems must be contextual: the truth-value assigned to a given proposition will depend on the context in which it is considered. The idea for their proof is the following (see Section \ref{sec:contextuality-in-QM} for detail). Imagine we have a set of $N$ binary observables $\mathcal{P}=\{ P_1, P_2, ..., P_N \}$ corresponding to yes-no questions about a quantum particle. Each $P_i$ is a Hermitian projection operator in a Hilbert space (in KS's paper a three dimensional one). As is well known, each $P_{i}$ is associated with a proposition about the quantum system. KS constructed a set of such operators with the following characteristics. First, there were several subsets of three commuting operators, such that one and only one of them were true for this set (i.e., they were orthogonal, and their sum was one). We can think of these subsets as a context, determined by the set of simultaneous propositions considered. These subsets had the additional feature that each $P_i \in \mathcal{P}$ appeared twice, one time for each of two possible contexts.  By constructing an appropriate set $\mathcal{P}$, KS showed that the structure of quantum observables and their corresponding contexts did not allow the consistent assignment of truth values for each $P_i$ that was the same for \emph{all~contexts}. Thus, in this sense, only \textit{contextual} hidden variable theories are compatible with the quantum formalism. Furthermore, this contextuality exists for all quantum systems that are complex enough (more specifically, it holds for any Hilbert space of dimension greater than two). 

Further study in hidden variable models led to the discovery of the so-called non-contextuality inequalities. These can be experimentally testable, opening an obvious field of research for discarding theories that deviate from experiments (and quantum theory). Examples of them are the KCBS inequalities in \cite{Klyachko-2008} and the GHZ inequalities in \cite{deBarros-Suppes-2000}. It was later shown that Bell and CHSH inequalities fall into this category. These inequalities' characteristic feature is that they put an upper bound on the correlations that a family of non-contextual hidden variable theories can model. Thus, an approach is non-contextual if the correlations predicted by it satisfy a specific bound. Since the correlations predicted by quantum theory do violate those inequalities, it is natural (and tempting) to say that quantum mechanics is contextual. Notice that this is a shift from the old quantum physics jargon, for which only hidden-variable theories could be considered as contextual or not. 

Furthermore, in the last decades, this quantum theory feature has attracted a lot of interest due to its potential role in quantum information processing tasks. Thus, instead of being considered a negative characteristic, nowadays, physicists seeking to develop quantum technologies, consider contextuality a positive feature of quantum theory itself, which can be quantified, measured, and used as a resource. In this work, we will follow the current jargon, and refer to the feature of the quantum formalism discovered by Kochen and Specker as \textit{quantum contextuality}. In other words, we will use expressions such as ``quantum mechanics is contextual,'' ``this theory (or state) has such amount of contextuality,'' and so on, to simply mean that outcomes of experiments are contextual.

There is yet another --less explored-- feature of quantum mechanics that justifies the modern jargon. Propositions about quantum systems are linked to concrete experimental settings, which are selected by the experimenter. If we prepare a quantum system in a particular state and consider a proposition in a given context, we find empirically that the result of an experiment might not be the same should we repeat the test with the same state, but with the given proposition considered in a different context. This is phenomenologically given, and it is independent of any interpretation. Furthermore, one might avoid speaking about states at all, and only refer to preparations and testable quantities of physical systems and their correlations in a theory-independent way; still, it would be meaningful to determine whether experiments display contextuality or not, and this could be checked by observing probability distributions and non-contextuality inequalities objectively. If a system shows contextual correlations, we refer to this feature by saying that the system is \textit{empirically contextual}. This notion of empirical contextuality is consistently defined, objectively testable, and it is model-independent (in the sense that they only assume very general features of probabilistic models).

Because of contextuality, one cannot represent quantum states with classical probabilities. Usually, one represents them by trace operators acting on a separable Hilbert space. But it seems possible to describe quantum states with extended probabilities. For example, the Wigner function takes a quantum state and transforms it into a classical phase space function. This function resembles a Kolmogorovian probability, but it may take negative values. Because it may be negative, it is considered a \emph{quasi-probabilities}. Most approaches to quasi-probabilities rely on an underlying theory (such as quantum mechanics) whose states and observables are mapped to a classical phase space in which the states take the form of quasi-probabilities (see for example \cite{Singer-Stulpe-PhaseSpaceRepresentations}). 

In this work, we take an alternative approach and focus on two aspects of quantum contextuality. First, we rely on the notions of signed measurable space and measurement context to give a formal definition of negative probabilities that is general enough to cover all cases of interest in quantum contextuality (and hopefully also outside of physics). Classical probabilistic models are shown to be particular cases of our formulation, which is general enough to include contextual models, such~as those coming from the quantum formalism. The approach presented here has many features in common with previous ones (see, for example, \cite{abramsky_sheaf-theoretic_2011,abramsky_operational_2014,de_barros_negative_2016}). Still, it relies more directly upon the notions of compatible random variables (for which a joint probability distribution exists), and thus, it provides a straightforward extension of Kolmogorov's approach. Our signed probabilities are constructed as no-signaling, meaning that the quasi-probability distribution associated with a random variable is context-independent. This particular feature is particularly relevant in physics, given that all physical theories satisfy this condition.

The other focus of this article is on quantum indistinguishability. In previous works, we have discussed the connection between particle and property indistinguishability as related to contexts \cite{de_barros_indistinguishability_2019}. Here we show that property indistinguishability leads to the no-signaling condition.  Since negative probabilities are necessary and sufficient for the description of no-signaling models, we argue that there is a connection between the principle of particle indistinguishability and negative probabilities. The assumption of indistinguishability for quantum particles leads to contextual and indistinguishable properties, which can, in turn, be naturally modeled using our definition of signed probabilities.

We organize this paper as follows. After reviewing elementary facts about contextuality in Section~\ref{sec:contextuality-in-QM}, in Section \ref{sec:Negative-Probabilities} we motivate and provide our definition of signed probabilities. In Section~\ref{sec:indistinguishability}, we discuss the connection between quantum indistinguishability, negative probabilities, and the non-signaling condition. Finally, in Section \ref{sec:conclusions}, we end with some final remarks and conclusions.

\section{Contextuality in Quantum Mechanics}\label{sec:contextuality-in-QM}

Context is a term that comes from linguistics, especially from semantics and pragmatics \cite{stalnaker_context_1999}. For instance, in semantics, the truth-value of an utterance or written text may depend on the other statements or sentences that precede or follow it. Take the written sentence: ``Alice sat by the bank to observe the people.''  Its truth-value varies depending on other comments that accompanied it: if it were preceded by ``The river was calming and beautiful,'' its meaning would differ from if it were preceded by ``The heist needed planning.'' For the case where ``river'' preceded the sentence,  ``bank'' likely refers to the bank side of a river, whereas for the ``heist'' case, ``bank'' refers to a financial institution. Though this is a case where meaning changes, there are other examples in linguistics where meaning does not change, but truth-value does. We can think of those as examples of context-dependency, or~contextuality, in linguistics \cite{de_barros_rationality_2019}. 

Contextuality, as conceptually discussed above, is a central concept in the foundations of quantum mechanics. It is also the main driving difficulty in defining properties for quantum particles or systems. So, let us examine how contextuality appears in quantum mechanics by discussing the famous Kochen-Specker theorem \cite{kochen_problem_1967}. Here we present a more straightforward proof involving only nine contexts \cite{cabello_simple_2013}.

We start with a four-dimensional Hilbert space, $\mathcal{H}$. According to the standard formalism of quantum mechanics, measurable properties are represented by Hermitian operators in $\mathcal{H}$ (known as observables). A quantum system is said to have a property if an experiment measuring it yields the same value all the time. In the formalism, this translates into having the system be in an eigenstate of the Hermitian operator. A particularly important subset of observables is projection operators, which correspond to 0- or 1-valued observables. We can think of these binary properties as truth-values: either the quantum system has the property (1), or it does not (0).  To distinguish between general properties and those associated with projection operators, we call the latter testable propositions, or, in~short, propositions. The distinction between testable propositions and properties is subtle and debated in the literature (see, e.g., \cite{randall1983properties,da2013modal}). Here we use the terminology that propositions are a particular type of observables, as discussed above.   

A vector in $\mathcal{H}$ uniquely determines a projection operator. For example, the vector $|1,0,0,0\rangle \in \mathcal{H}$ corresponding to the column matrix with the first component as one and the others as zero determines the projector operator $\hat{P}_{1,0,0,0} \equiv |1,0,0,0\rangle \langle 1,0,0,0|$. Let us consider now the following set of equations. 
\begin{align}
\hat{P}_{0,0,0,1}+\hat{P}_{0,0,1,0}+\hat{P}_{1,1,0,0}+\hat{P}_{1,-1,0,0} & =1,\label{eq:cabello-1st}\\
\hat{P}_{0,0,0,1}+\hat{P}_{0,1,0,0}+\hat{P}_{1,0,1,0}+\hat{P}_{1,0,-1,0} & =1,
\label{eq:cabello-2nd} \\
\hat{P}_{1,-1,1,-1}+\hat{P}_{1,-1,-1,1}+\hat{P}_{1,1,0,0}+\hat{P}_{0,0,1,1} & =1,\\
\hat{P}_{1,-1,1,-1}+\hat{P}_{1,1,1,1}+\hat{P}_{1,0,-1,0}+\hat{P}_{0,1,0,-1} & =1,\\
\hat{P}_{0,0,1,0}+\hat{P}_{0,1,0,0}+\hat{P}_{1,0,0,1}+\hat{P}_{1,0,0,-1} & =1,\\
\hat{P}_{1,-1,-1,1}+\hat{P}_{1,1,1,1}+\hat{P}_{1,0,0,-1}+\hat{P}_{0,1,-1,0} & =1,\\
\hat{P}_{1,1,-1,1}+\hat{P}_{1,1,1,-1}+\hat{P}_{1,-1,0,0}+\hat{P}_{0,0,1,1} & =1,\\
\hat{P}_{1,1,-1,1}+\hat{P}_{-1,1,1,1}+\hat{P}_{1,0,1,0}+\hat{P}_{0,1,0,-1} & =1,\\
\hat{P}_{1,1,1,-1}+\hat{P}_{-1,1,1,1}+\hat{P}_{1,0,0,1}+\hat{P}_{0,1,-1,0} & =1.\label{eq:cabello-last}
\end{align} 
Each equation above is numerically equal to one because all the vectors in each line form a complete and orthonormal basis for  $\mathcal{H}$. This means that, for each Equations (\ref{eq:cabello-1st})--(\ref{eq:cabello-last}), we have four true-false properties that are compatible, complete, and mutually exclusive. Therefore exactly one of them must be true, and the others zero, which means they all add to one. 

An issue may be evident to some readers about (\ref{eq:cabello-1st})--(\ref{eq:cabello-last}): if we assign to each property a truth-value of zero or one we reach a contradiction. To see this contradiction, consider that each property $\hat{P}_i$ appears on the left hand side of   (\ref{eq:cabello-1st})--(\ref{eq:cabello-last}) twice. Since $2\hat{P}_i$ is an even number, it follows that the sum of all the terms on the left-hand side of (\ref{eq:cabello-1st})--(\ref{eq:cabello-last}) must be even. However, we add the right-hand side of (\ref{eq:cabello-1st})--(\ref{eq:cabello-last}) we total nine, clearly not an even number, which is a mathematical contradiction.
 
The mathematical contradiction is a result of assuming that the truth-value of a property $\hat{P}_i$ is the same when it is co-measured with different properties. For example, $\hat{P}_{0,0,0,1}$ shows up in (\ref{eq:cabello-1st}) but also in (\ref{eq:cabello-2nd}). However, the co-measured variables to $\hat{P}_{0,0,0,1}$ in (\ref{eq:cabello-1st}) are all different from the ones in~(\ref{eq:cabello-2nd}). In~the example above, therefore, we have nine contexts, and each property shows up in exactly two of those contexts. If we allow, for example, $\hat{P}_{0,0,0,1}$ to have a different truth-value when co-measured with $\hat{P}_{0,0,1,0}$, $\hat{P}_{1,1,0,0}$, and $\hat{P}_{1,-1,0,0}$ (call it Context 1) from when it is co-measured with $\hat{P}_{0,0,0,1}$, $\hat{P}_{0,1,0,0}$,  $\hat{P}_{1,0,1,0}$, and $\hat{P}_{1,0,-1,0}$ (Context 2), we reach no contradiction. It is in this sense that contextuality is claimed for quantum observables: the truth-value of a property varies with its context determined by the collection of co-measured properties. 

The above example has some intriguing features. First, it is state-independent. This feature means that it does not matter how we prepare the quantum system; if we try to measure the properties on (\ref{eq:cabello-1st})--(\ref{eq:cabello-last}), they will change from context to context. Therefore contextuality is a property of the quantum-operator algebra. Second, what the KS theorem shows is a \emph{logical contradiction} that arises from a context-independence assumption. This means that we do not need to involve probabilities in proving the contextuality of quantum properties. 

However, probabilities are a fundamental aspect of quantum theory, and perhaps of any empirical theory. So, how could we formulate the KS theorem in terms of probability theory? The hint can be found on \cite{deBarros-Suppes-2000}: logical inconsistencies are but a special case of probability one events when a joint probability distribution does not exist that describes the outcomes of the experiments. To see this, let us consider the example of four two-valued properties, $A$, $A'$, $B$, and $B'$, who can only be observed in the following pairwise experimental arrangements: $A$ with $B$; $A$ with $B'$; $A'$ with $B$; and $A'$ with $B'$. If~we assume that those properties are context-independent, then the combination of their values defined~by 
\begin{equation}
\label{eq:S-defined}
    S=AB+AB'+A'B-A'B'
\end{equation}
is always a number equal or less than two. The reader can verify the previous statement for all possible combinations, but as an example, if $A=1$, $A'=-1$, $B=1$, and $B'=1$, $S=1+1-1+1=2$. Since any combination of $A$, $A'$, $B$, and $B'$ yields a value of $S$ that is 2 or less, it follows that convex combinations of $S$ imply that 
\begin{equation}
    \langle S \rangle \leq 2,
    \label{eq:S-inequality}
\end{equation}
where we are using the fact that the mean value of $S$, denoted $\langle S \rangle$, is a convex combination of each of its possible values. It follows, from (\ref{eq:S-inequality}) that if $S>2$, there is no convex combination of the \emph{logical} context-independent possibilities that yields the expected value of $S$. In other words, it is not possible to assign probabilities to  the possible combinations of values of $A$, $A'$, $B$, and $B'$ consistent with $\langle S \rangle > 2$. This is why a joint probability distribution for $A$, $A'$, $B$, and $B'$ does not exist, although, of course, marginal probabilities do, since we can use the data tables to, say, compute the value of $\langle AB \rangle $.

We should point out that (\ref{eq:S-inequality}) is one of the CHSH inequalities \cite{clauser_proposed_1969}. By itself, as we saw above, a violation of (\ref{eq:S-inequality}) is sufficient to establish the non-existence of a joint probability distribution or contextuality for the observables in question. However, other inequalities need to be added to (\ref{eq:S-inequality}) to form a set of necessary and sufficient conditions for the contextuality of properties. 

The CHSH inequalities \cite{clauser_proposed_1969} are related to Bell's inequalities \cite{bell_problem_1966}, and they can be used to show that quantum mechanics is a non-locally contextual theory, or simply non-local. This is done by starting with two spin-$1/2$ particles, $A$ and $B$, in an entangled state 
\begin{equation}
    |\psi\rangle = \frac{1}{\sqrt{2}}\left( |+-\rangle - |-+\rangle \right),
    \label{eq:Bell-entangled}
\end{equation}
where $|+-\rangle$ is the state where particle $A$ has spin $+1/2$ and $B$ spin $-1/2$ and $|-+\rangle$ the other way around. It is easy to prove from (\ref{eq:Bell-entangled}) that the joint expectation of two spin measurements in directions $\theta_1$ for $A$ and $\theta_2$ for $B$ yield the following correlation: 
\begin{equation}
    \label{eq:Tsirelson-bound}
    E(\theta_1 , \theta_2 ) = \sin (\theta_1 -\theta_2).
\end{equation} 
The reader can  verify that for the combinations of measuring the spin of $A$ at 0\degree  and $45\degree$ and $B$ at $22.5\degree$ and $67.5\degree$, $\langle E \rangle = 2\sqrt{2}>2$, which violates (\ref{eq:S-inequality}). So, quantum mechanics is not only contextual, but~its contextuality manifests for observers that may be far apart from each other, such as the case of the two-particle example above. Contextuality appears in quantum mechanics from the structure of the Hilbert space and that it is present even for systems whose properties are space-like separated. This~contextuality presents difficulties to the concept of property in quantum mechanics, as they would depend on the experimenter's choice of a measurement apparatus, as discussed above.  

To summarize, in this section, we discussed the idea of contextuality both from an intuitive and formal perspective. We saw that contextuality is the impossibility of consistently assigning truth-values to the same testable proposition in different contexts. Equivalently, a similar assertion holds for observables: it is impossible to assign non-contextual values to all possible observables if some minimal functionality conditions are to be considered \cite{kochen_problem_1967}. Alternatively, one can interpret contextuality as the proposition (or observable) changing from one context to another. These observations lead to a subtle (but fundamental) problem: do propositions (or observables) retain their identity when considered in different contexts? Let us be more explicit about this. In the scenario described above, consider the contexts $AB$ and $AB'$. What is the status of observable $A$ in contexts $AB$ and $AB'$? Let us denote $A_{B}$ and $A_{B'}$ to the observable $A$ considered in contexts $AB$ and $AB'$, respectively. Usually, since quantum systems obey the no-signal condition, physicists tend to identify $A_{B}$ and $A_{B'}$ (i.e., $A_{B}=A_{B'}$). However, this assumption is not trivial at all and has indeed been criticized. In some fields of research, $A_{B}$ and $A_{B'}$ may not have the same distribution (as is the case in signaling theories) and, even if they have the same content, it should be dubious to identify them. Some authors have proposed that $A_{B}$ and $A_{B'}$ should be considered different whenever a system manifests a strong degree of contextuality \cite{dzhafarov_contextuality-by-default_2017,dzhafarov_contextuality-by-default_2019}. In previous works \cite{de_barros_contextuality_2017,de_barros_indistinguishability_2019}, we have proposed an alternative solution to the dichotomy $A_{B}=A_{B'}$ vs $A_{B}\neq A_{B'}$. Using a formal framework that allows dealing with collections of indistinguishable objects (see Section \ref{sec:indistinguishability} of this work), we have proposed that $A_{B}$ and $A_{B'}$ can be thought of as indistinguishable (denoted by $A_{B}\equiv A_{B'}$). This point of view allows us to connect with contextuality one of the most fundamental features of quantum theory: quantum systems of the same kind are indistinguishable.  More specifically, we show in \cite{de_barros_contextuality_2017,de_barros_indistinguishability_2019} that the indistinguishability of particles leads to the indistinguishability of propositions and that this, in turn, gives place to contextuality. In~the rest of this work, we elaborate on these ideas further and show a strong connection between the indistinguishability of testable propositions (or observables) and negative probabilities. To do this, we~must first introduce a definition of negative probabilities that is useful for our purposes and general enough to cover all physical models of interest.

\section{Negative Probabilities}\label{sec:Negative-Probabilities}

Negative Probabilities (NP) have a long tradition in physics and find applications in different branches of quantum physics \cite{Hillery-PhysicsReports}. NP appeared in physics early in the 20th century in quantum mechanics, for example, in connection to the Klein-Gordon equation or Wigner's paper on the classical approximations for quantum statistical mechanics \cite{wigner_quantum_1932}. However, NP were considered an undesirable side effect of a defective model or theory. As such, theories yielding NP were discarded as having no physical interest. The first physicist to take NP seriously was Dirac, who used them as the basis for his interpretation of the theory of photons \cite{Dirac-1942}. They also were discussed by Feynman, who thought they were a promising concept but could not find any use for them \cite{Feynman-1987}. Nevertheless, their study helped understand the connection and differences between quantum and classical systems. In some fields --as is the case in quantum optics-- they have even become a tool of everyday use  \cite{Glauber-Negative}. Furthermore,  they~form the basis of many contextuality measures \cite{de_barros_measuring_2015,kujala2019measures} and serve to characterize quantumness of states and theories \cite{abramsky_sheaf-theoretic_2011}. Recent studies aim to understand the differences between the correlations originated in quantum theory and those that come from other plausible no-signaling generalized probabilistic models \cite{SimulatingNon-signaling}. In this setting, negative probabilities are used to characterize different features of quantum mechanics \cite{Spekkens-NegativityandContextuality-PRL,Singer-Stulpe-PhaseSpaceRepresentations}. Nowadays, NP have become a fundamental tool in quantum information theory and the development of quantum technologies. In particular, they play a significant role in the problem of quantum state estimation \cite{QuantumTomography-DiscreteWigner}, the determination of quantum correlations and classicality of quantum states \cite{Cormik-ClassicalityDiscreteWigner}, and the study of quantum computers' speed-up \cite{Veitch_2012,Galvao-WignerSpeedUp}. 

In our discussion of NP, let us start with Wigner's work. In his 1932 paper \cite{wigner_quantum_1932}, Wigner asked the following question: if we have an ensemble of $N$ classical particles, what types of corrections would we have to introduce to their phase-space probability distributions such that their statistics coincided with the quantum one. For this purpose, he constructed what is now known as the Wigner distribution, given by
\begin{equation}
    W(\textbf{r},\textbf{p}) = \frac{1}{(2\pi)^3}\int \psi^{*}\left(\textbf{r}+\frac{\hbar}{2}\textbf{s}\right)
    \psi\left(\textbf{r}-\frac{\hbar}{2}\textbf{s}\right)
    e^{i\textbf{p}\cdot \textbf{s}} d^3\textbf{s},
\end{equation}
where $\textbf{r}$ and $\textbf{p}$ are the position and momentum, and $\textbf{s}$ is an integration variable. A similar definition holds for arbitrary pairs of conjugate variables. It is easy to see that $W$ behaves similarly to a joint probability distribution, in the sense that if we integrate $W$ on either $\textbf{r}$ or $\textbf{p}$ we get the marginal probability distributions. For example,
\begin{equation}
    \int W(\textbf{r},\textbf{p}) d^3 \textbf{p} = |\psi(\textbf{r})|^2. 
\end{equation}
However, as Wigner pointed out, $W$ is not a proper joint probability distribution, as it can take negative values. For example, for the ground state of the harmonic oscillator, $W$ is non-negative, but for the first excited state, it is negative in some regions of the phase space \cite{suppes_probability_1961}. After Wigner, Dirac \cite{dirac_bakerian_1942} used negative probabilities to try to solve the problem of infinities in quantum field theory. In his theory, negative probabilities were nothing more than an accounting tool for computing (non-negative) observable probabilities, and carried the same interpretation as the statement ``having negative three~apples.'' This was similar to the interpretation suggested by Feynman in his article on negative probabilities \cite{Feynman-1987}. For a review of the history of negative probabilities in physics, the interested reader is referred to \cite{muckenheim_review_1986}. More recently, negative probabilities have been used in foundations of quantum mechanics, and the interested reader is referred to references \cite{de_barros_negative_2016,oas_exploring_2014,SimulatingNon-signaling} and references therein. For~possible interpretations of negative probabilities that are not based on a pragmatic bookkeeping, readers are referred to \cite{abramsky_operational_2014,burgin_interpretations_2010,khrennikov_interpretations_2009,szekely_half_2005,machado_fractional_2013}.  

What are negative probabilities? Let us start with the standard probability theory. The currently accepted axioms for probability were laid down by Kolmogorov \cite{kolmogorov_foundations_1956}. In his axioms, we start with a sample set $\Omega$, which we can think of as possible states of the system of interest. For example, if we are interested in a die's outcomes, $\Omega$ could be the set $\{ 1,2,3,4,5,6\}$. We could, in principle, talk about the probabilities of the members of $\Omega$. Still, Kolmogorov recognized that, in probability theory, we want to refer to logical combinations of possible states. To do so, he associated with $\Omega$ a $\sigma$-algebra $\mathcal{F}$ of its elements. Once we have $\Omega$ and $\mathcal{F}$, he define the probability $p$ as a non-negative real-valued function $p:\mathcal{F}\rightarrow [0,1]$ satisfying the following properties. 
\begin{description}
\item[K1.] $p(\Omega)=1$
\item[K2.] For every denumerable and disjoint family $\{A_{i}\}_{i\in\mathbb{N}}$, $p(\bigcup A_{i})=\sum_{i} p(A_{i})$.
\end{description}

It is easy to see, for simple examples, that Kolmogorov's definition captures the essence of probabilities first put forth by Pascal and then developed throughout the centuries (for a wonderful historical account of probability theory, see  \cite{galavotti_philosophical_2005}.). 

However, as we saw in Section \ref{sec:contextuality-in-QM}, it is not always possible to have a joint probability distribution that accounts for all experimental outcomes. There are different ways to approach this lack of a joint. One possibility is to notice that the algebra of observables is not Boolean, but follows a lattice structure that does not allow for certain Boolean operations (for example, the complement of a property may not exist) \cite{foulis_operational_1972}. This is the quantum logic approach, and one could try to create a probability calculus over lattices, and not Boolean algebras. Of course, one such probability calculus is the Hilbert space formalism. Another approach could be to modify Kolmogorov's definition to allow for a new probability function, say $p^*$, to exist. For example, we could change K2 from an equality to an inequality, as is the case for upper and lower probabilities \cite{suppes_existence_1991, de_barros_probabilistic_2010,holik_discussion_2014}. Another possibility is to keep the algebra intact, as well as K1 and K2, but change the requirement that $p$ is non-negative, i.e., to allow for negative probabilities.  

What are the axioms for negative probabilities?  To give a straightforward description based on measure theory (obtaining thus a canonical generalization of Kolmogorov's approach), we rely on the notion of compatible random variables and signed measure spaces. In the rest of this section, we will try to motivate and write down a definition for negative probabilities in the spirit of Kolmogorov.

Let us start with a definition of random variables. 
\begin{Definition}
\label{def:random-variables}
Let $(\Omega,\mathcal{F},p)$ be a probability space, and let $(M,\mathcal{M})$ be a Borel space with elements of $M$ being real numbers, i.e., $\mathcal{M}$ is a $\sigma$-algebra over $M$. A (real-valued) random variable $\mathbf{R}$ is a measurable function $\mathbf{R}:\Omega \rightarrow M$, i.e., for all $m\in \mathcal{M}$, $\mathbf{R}^{-1}(m)\in \mathcal{F}$. 
\end{Definition}
Though the above definition may seem complicated, it is intuitive. What it says is that we can associate to partitions of the sample space $\Omega$ a particular real number. A simple example is the game of craps. Imagine we throw two dice and record their outcomes. A sample space for this example is ${(1,1),(1,2),\ldots,(6,6)}$, where each ordered pair corresponds to an outcome for each die. In a game of craps, often, what matters is the sum of the values and not the individual outcomes. For example, rolling a \textit{seven out}, a sometimes desired outcome, is the result of one of the following outcomes: $(1,6)$, $(2,5)$, $(3,4)$, $(4,3)$, $(5,2)$, or $(6,1)$. A random variable yielding the sum of the thrown dice would associate to all those outcomes the value $7$. As defined, random variables are a way to model outcomes of experiments or observations that are stochastic, i.e., that have certain randomness associated with~them. 

If we look back at our examples in Section \ref{sec:contextuality-in-QM}, we can see that random variables may express contextuality. For example, let us consider the four two-valued properties $A$, $A'$, $B$, and $B'$. Since they could be used to describe yes/no properties, let us think of each of them as a $\pm 1$-valued random variables in a given a probability space $(\Omega,\mathcal{F},p)$, e.g., $\mathbf{A}:\Omega \rightarrow {1,-1}$. In terms of random variables,~(\ref{eq:S-defined})~would be rewritten simply as 
\begin{equation}
    \label{eq:S-random-variables}
    \mathbf{S} = \mathbf{AB}+\mathbf{A}\mathbf{B}'+\mathbf{A}'\mathbf{B}-\mathbf{A}'\mathbf{B}'.
\end{equation}
Since it follows from standard probability theory that 
\begin{equation}
\label{eq:CHSH}
\langle \mathbf{S}\rangle = \langle \mathbf{AB} \rangle+\langle\mathbf{A}\mathbf{B}'\rangle+\langle \mathbf{A}'\mathbf{B}\rangle -\langle \mathbf{A}'\mathbf{B}'\rangle \leq 2,
\end{equation} any violation of this inequality would imply that no (standard) probability space exists that allow for the correlations observed in those random variables. Equation (\ref{eq:CHSH}) is one of the well-known CHSH inequalities, which are necessary and sufficient conditions for the existence of a joint probability distribution \cite{clauser_experimental_1974,clauser_proposed_1969}.  However, for this example, it is trivial to construct four different probability spaces for each experimental situation, i.e. $A$ and $B$, $A'$ and $B$, $A$ and $B'$, and $A'$ and $B'$. The impossibility is to find a single probability space that yields all four correlations that are experimentally observed in quantum theory. And this is how random variables can help us define negative probabilities. We~can relax the non-negativity assumption as long as we guarantee that all observable properties do not result in negative probabilities (we point out that with weak measures, negative probabilities may be ``observable,'' but we will not discuss this here; readers are referred to \cite{sokolovski_weak_2007,hosoya_strange_2010}). This motivates the following definitions. 
\begin{Definition}
\label{def:signed-measure}
Let $\Omega$ be a sample space and $\mathcal{F}$ a $\sigma$-algebra over $\Omega$. A signed measure is a function $\mu : \mathcal{F}\rightarrow \mathbb{R}$ such that 
\begin{equation}
    \mu (\emptyset)= 0
\end{equation}
and for every denumerable and disjoint family $\{A_{i}\}_{i\in\mathbb{N}}$
\begin{equation}
    \mu (\bigcup_{i} A_{i})= \sum_{i}\mu (A_{i}) 
\end{equation}
The triple $(\Omega,\mathcal{F},\mu)$ is called a signed measure space \cite{halmos_measure_1974}. 
\end{Definition}
Signed measure spaces expand the idea of measures (not probabilities), to the negative domain. However, it should be clear to the reader that signed measures are a generalization of probability measures, one we will use to define negative probabilities. 
\begin{Definition}
\label{def:extended-random-variables}
Let $(\Omega,\mathcal{F},\mu)$ be a signed measure space, and let $(M,\mathcal{M})$ be a Borel space with elements of $M$ being real numbers, i.e., $\mathcal{M}$ is a $\sigma$-algebra over $M$. A (real-valued) extended random variable $\mathbf{R}^*$ is a measurable function $\mathbf{R}^*:\Omega \rightarrow M$, i.e., for all $m\in \mathcal{M}$, $(\mathbf{R}^*)^{-1}(m)\in \mathcal{F}$. 
\end{Definition}
Notice that extended random variables are not at all equivalent to random variables, except in special cases when $\mu$ is a probability measure. 
\begin{Definition}
\label{def:NEWcontext}
Let $\{R^*_i\}$, $i=1,\dots, n$, be a collection of extended random variables defined on a signed measure space $(\Omega,\mathcal{F},\mu)$. A \emph{ $\mu$-induced context} is a subset  $C^{\mu}_j=\{R^*_{k}\}_{k\in N_j}$, $N_j \subset \{ 1,\ldots, n\}$, for which there exists a sub-$\sigma$-algebra $\mathcal{F}_{j}$ of $\mathcal{F}$ such that, by defining $p^{\mu}_{j}(F):=\mu(F)$ for all $F\in\mathcal{F}_{j}$, the triad $(\Omega,\mathcal{F}_{j},p^{\mu}_{j})$ becomes a probability space, and $R^*_{i_{k}}$ is a random variable with respect to it, for all $k\in \{1,...,n_{j}\}$.
\end{Definition}

Some observations are in order. First, the notion of context given by Definition \ref{def:NEWcontext} depends on the chosen measure $\mu$. Since we are grounding our definitions on measure theory, the available mathematical tools are a set $\Omega$, a collection $\mathcal{F}$ of subsets of it (forming a Boolean algebra), and a signed measure $\mu$. The dependence on $\mu$ makes our definition of context \textit{measure dependent}. We aim to represent each possible state of the system under study by a normalized signed measure. A concrete probabilistic model for a system is determined when all its possible states are specified. Once this is done, the contexts of the theory can be unambiguously determined as follows. We denote by $\mathcal{S}$ to the collection of all possible states of a system, described as signed measurable spaces. In order to obtain a consistent theory (such as a classical or quantum probability theory), we assume that all states have associated the same outcome set $\Omega$ and the same $\sigma$-algebra $\mathcal{F}$ and that they are normalized. It is useful to put this in terms of a definition.

\begin{Definition}
Let $\Omega$ be a set and $\mathcal{F}$ a $\sigma$-algebra of subsets of $\Omega$. A family of signed probabilistic models for $(\Omega,\mathcal{F})$ is a collection $\mathcal{S}_{(\Omega,\mathcal{F})}$ of signed measures on $(\Omega,\mathcal{F})$ such that, for all $\mu\in \mathcal{S}_{(\Omega,\mathcal{F})}$, $\mu(\Omega)=1$. Any $\mu\in \mathcal{S}_{(\Omega,\mathcal{F})}$ is called a state of the model.
\end{Definition}

The above definition is analogous to that of states in a classical probabilistic model, the sole difference being that we allow the states to take negative values. In order to describe the observables of physical theories, we need each extended random variable to be consistently defined with regard to all possible states $\mathcal{S}_{(\Omega,\mathcal{F})}$ in the following sense. Considered as a function $R^*_i:\Omega\longrightarrow\mathbb{R}$, we must have that each extended random variable must satisfy $(R^*_i)^{-1}(\Delta)\in\mathcal{F}$, for every Borel set $\Delta\subseteq\mathbb{R}$ (this means that the $R^*_i$'s are measurable functions with regard to all possible $\mu\in\mathcal{S}_{(\Omega,\mathcal{F})}$). This condition grants that the extended random variables are well defined for all $\mu\in\mathcal{S}_{(\Omega,\mathcal{F})}$. With these definitions, we are ready to provide a state-independent definition of context.

\begin{Definition}
\label{def:GlobalContext}
Consider a family of signed probability models $\mathcal{S}_{(\Omega,\mathcal{F})}$. Let $\{R^*_i\}$, $i=1,\dots, n$, be a collection of  extended random variables defined on $\mathcal{S}_{(\Omega,\mathcal{F})}$. A general context is a subset  $C_j=\{R^*_{k}\}_{k\in N_j}$, $N_j \subset \{ 1,\ldots, n\}$ of those extended random variables, for which there exists a sub-$\sigma$-algebra $\mathcal{F}_{j}$ of $\mathcal{F}$ satisfying that, for all $\mu\in\mathcal{S}$, by~defining $p^{\mu}_{j}(F):=\mu(F)$ for all $F\in\mathcal{F}_{j}$, the triad $(\Omega,\mathcal{F}_{j},p^{\mu}_{j})$ becomes a probability space, and $R^*_{i_{k}}$ is a random variable with respect to it, for all $k\in \{1,...,n_{j}\}$.
\end{Definition}
Using the definition of general context, we can naturally introduce the notion of \emph{signed probability space} as follows.
\begin{Definition}
\label{def:SignedProba}
A signed probability space, also called here negative probability space, is a signed measure space $(\Omega,\mathcal{F},\mu)$ endowed with a non-empty set of contexts $C=\{C^{\mu}_{j}\}$ (in the sense of Definition \ref{def:NEWcontext}), such that $\mu(\Omega)=1$. The measure $\mu$ in this space is a \emph{signed probability} or negative probability.
\end{Definition}
In other words, a signed probability space is a signed measure space for which there exist contexts, and these contexts give place to well defined probabilistic scenarios. 
\begin{Proposition}
If a state $\mu\in\mathcal{S}_{(\Omega,\mathcal{F})}$ of an extended probabilistic model admits a non-empty set of contexts, then, it defines a signed probability space. 
\end{Proposition}
\begin{proof}
If $\mu\in\mathcal{S}_{(\Omega,\mathcal{F})}$ is a state, then, $\mu$ is a signed measure on $(\Omega,\mathcal{F})$ such that $\mu(\Omega)=1$. Thus, the existence of a non empty family of contexts for $(\Omega,\mathcal{F},\mu)$, makes it satisfy Definition \ref{def:SignedProba}.
\end{proof}

After the above Definitions, it is important to make the following remarks.
\begin{Proposition}
If $(\Omega,\mathcal{F},p)$ is a probability space, then it is also a signed probability space.
\end{Proposition}
\begin{proof}
Any $(\Omega,\mathcal{F},p)$ satisfying Kolmogorov's axioms also satisfies the axioms of signed measure in Definition \ref{def:signed-measure}. Given that $p$ is normalized, it is also a state with respect to the pair $(\Omega,\mathcal{F})$. Any~collection of random variables defined on $(\Omega,\mathcal{F},p)$, induces a context satisfying Definition \ref{def:NEWcontext} (by taking sub-$\sigma$-algebra as $\mathcal{F}$ itself). Thus, the states of classical probabilistic systems can be described as a particular case of signed probabilities.
\end{proof}
\corres{}
%\begin{Proposition}
%For each quantum state $\phi\in \mathcal{H}$  Using the isomorphism between quantum states and Wigner functions, we obtain that each quantum state can be considered as a negative probability that satisfies definition \ref{def:SignedProba}. 
%\end{Proposition}

The states of the extended probability model of quantum theory are just the quantum states' images under the Wigner transform.  Any context of a quantum system---understood in the usual sense of a family of commuting observables---can be described in our approach by a collection of extended random variables.

Definitions \ref{def:NEWcontext}, \ref{def:GlobalContext}, and \ref{def:SignedProba} are inspired in the following properties of the Wigner distribution function. For simplicity, suppose that we
have a phase space
\mbox{$\Omega=\{(x,p)\in\mathbb{R}\times\mathbb{R}\}=\Omega_{1}\times\Omega_{2}$} (i.e., we~are taking $\Omega_{1}=\mathbb{R}=\Omega_{2}$).
Let $\mathcal{F}$ be the collection of Borel subsets of $\Omega$.
Then, we have that the quasi-probability of obtaining a system in
the set $F\in\mathcal{F}$ is given by
\mbox{$\mu(F):=\int\int_{F}W(x,p)dxdp$}, where $W(x,p)$ is the Wigner
distribution function. Indeed, this distribution defines a normalized signed
measurable space $(\Omega,\mathcal{F},\mu)$. To obtain the marginal
measures, we~must do as follows. Let~$\mathcal{F}_{1}$ be the subalgebra of $\mathcal{F}$ formed by all elements of the form $\Delta\times\Omega_{2}$, where $\Delta$ ranges over any possible Borel set of the real line. Define $W(x):=\int_{\Omega_{2}} W(x,p)dp$ and \mbox{$p^{\mu}_{1}(\Delta\times\Omega_{2}):=\int_{\Delta}W(x)dx=\int_{\Delta}\int_{\Omega_{2}}W(x,p)dxdp=\mu(\Delta\times\Omega_{2})$}. While $\mu$ is not in general a
positive measure, $p_{1}$ always is, and $(\Omega,\mathcal{F}_{1},p^{\mu}_{1})$ is indeed Kolmogorovian. It also coincides numerically with the probabilities for position context computed from the quantum formalism. A similar Kolmogorovian measure $(\Omega,\mathcal{F}_{2},p^{\mu}_{2})$ can be obtained in an analogous way for the momentum context. Further comments are in order:

\begin{itemize}[leftmargin=2.2em,labelsep=5mm]
\item Suppose that a random variable belongs to two different general contexts $C_{i}$ and $C_{j}$ (according to Definition \ref{def:GlobalContext}). For each $\mu\in\mathcal{S}$, the condition $p^{\mu}_{j}(F):=\mu(F)$ in Definition \ref{def:GlobalContext} implies that $p^{\mu}_{i}(F)=\mu(F)=p^{\mu}_{j}(F)$, for all events $F$ associated to this random variable. In other words, the~probability of a proposition is independent of the context in which it is tested. This implies that the probability distribution assigned to an observable will be independent of the other observables with which it is co-measured. This condition is nothing but the generalized version of the \emph{no-signaling condition} in physics (we will further discuss this below). It means that the probability of a given event (or more generally, the probability distribution of a given random variable) will not depend on the context in which it is considered. Thus, according to Definition \ref{def:GlobalContext}, all negative probabilities that we consider satisfy the no-signaling condition. 
\item In Definition \ref{def:GlobalContext}, for each $\mu$, all measurable functions defined over the probability space $(\Omega,\mathcal{F}_{j},p^{\mu}_{j})$ define legitimate observables
in the classical sense. These observables are all compatible. It is
in this sense that the $C_{j}$'s define contexts. If we mix an
observable from context $i$ with other taken from context $j$, there
is no reason to assume that there will exist a joint (Kolmogorovian) probability
distribution for them, because $\mu$ is not necessarily positive
definite. For example, the~proposition ``the observable $f_{i}$ (taken
from context $C_{i}$) possesses its value in the interval
$\Delta\in\mathcal{F}_{i}$ and the observable $g_{j}$ (from context
$j$) possesses its value in the set $\Gamma\in\mathcal{F}_{j}$", has
a quasi-probability given by $\mu(\Delta\times\Gamma)$. These observables are not necessarily compatible because, by construction, we allow this quantity to be negative. Being negative, this probability cannot be observed in any measurement context.
\end{itemize}

Each  context represents a real empirical scenario, where probabilities and observable  quantities are suitably defined. In general, given a set of random variables, it is not necessarily true that a joint probability distribution (understood in the Kolmogorovian sense) exists for all variables. However, for random variables describing physical measurements in different contexts, a negative probability distribution can always be constructed. Definition \ref{def:SignedProba} includes those cases. 

%In order to stay closer to more concrete physical examples (to fix ideas, think about the Wigner function), it is useful to also propose the following definition.
%\begin{Definition}
%\label{def:context2}
%Let $(\Omega,\mathcal{F},\mu)$ be a signed probability space. A canonical decomposition of $(\Omega,\mathcal{F},\mu)$, is a collection of probability spaces $(\Omega_{j},\mathcal{F}_{j},p_{j})$ associated to contexts, such that $\Omega\cong\Omega_{1}\times\Omega_{2}\times...\times\Omega_{N}$, in such a way that $p_{j}(\Delta_{j})=\mu(\Omega_{1}\times\Omega_{2}\times...\times\Delta_{j}\times...\times\Omega_{N}))$, for all $\Delta_{j}\in\mathcal{F}_{j}$.
%\end{Definition}

A typical practical situation is the following. Suppose that a collection of contexts $\{C_{j}\}$ is given and that there is more than one signed probability space in which those contexts are defined. Among all possible signed probability spaces compatible with a family of contexts, which one should we chose? To help us understand this question, we should define compatible signed probability spaces.

\begin{Definition}
A family of signed probability spaces is compatible if their collection of contexts is the same. 
\end{Definition}
Given a family of contexts $F=\{C_{j}\}$, call $\mathcal{S}(F)$ the maximal set of compatible signed probability spaces that have $F$ as its collection of contexts. Which signed probability space should we take among all possible in $\mathcal{S}(F)$? The problem of the existence of a ``minimal one'' is subtle and will be treated elsewhere. Instead, we give here the following definition, which is useful in many circumstances. We~also restrict to finite sets in order to simplify the analysis.

\begin{Definition}
\label{def:negative-probabilities}
Let $\mathbf{\Omega}_i =(\Omega_i,\mathcal{F}_i,\mu_i)$, $i\in I$, be a compatible collection of signed probability spaces. For~each $\mathbf{\Omega}_i$, let  $M_i = \sum_{\omega\in\Omega_i}|\mu_i(\omega)|$. Then $\mathbf{\Omega}_k$ is a minimal signed (or negative) probability space if \mbox{$M_k=\min \{M_i|i\in I\}$} when it exists.
\end{Definition}

From now on, we will use the notation $p^*$ for negative probabilities, $p$ for regular probabilities, and $\mu$ for measures that are not necessarily probabilities (signed or not). With this notation in mind, we can write the following results \cite{de_barros_negative_2016}.
\begin{Proposition}
\label{prop:no-signal-NP}
Let $\mathbf{\Omega}=(\Omega,\mathcal{F},p^*)$ be a minimum signed probability space. If $M=\sum_{\omega\in\Omega}|p^*(\omega)|=1$, then  $\mathbf{\Omega}$ is also a  probability space. Alternatively, if $\mathbf{\Omega}$ is a probability space, then it is also a minimum signed probability space, with $M=1$.
\end{Proposition}
\begin{proof}
Since, by Definition \ref{def:negative-probabilities}, we have $\sum_{\omega\in\Omega}p^*(\omega)=1$, it follows that $\sum_{\omega\in\Omega}|p^*(\omega)|=1$ implies $p^*(\omega)$ is non-negative for all $\omega\in\Omega$. Given that negative probabilities satisfy all of Kolmogorov's axioms except the non-negativity one, it follows that $p^*$ is a probability, if $M=1$. Alternatively, for~non-negative $p^*$ that add to one, it is immediate that the sum of their absolute value also add to one.  See reference \cite{de_barros_negative_2016} for details.
\end{proof}
The above Proposition suggests that the L1 norm plays an essential role in whether a probability distribution exists or not for a set of correlations and random variables. This motivates the following~definition. 
\begin{Definition}
\label{def:contextuality-index}
Let $\mathbf{\Omega}=(\Omega,\mathcal{F},p^*)$ be a minimal signed probability space. The quantity $\delta$, defined as $\delta = \sum_{\omega\in\Omega}|p^*(\omega)|-1 $ is called the contextuality index of $\mathbf{\Omega}$ or, in short, contextuality index.
\end{Definition}
The contextuality index provides a measure of contextuality for a set of experimental outcomes associated to observations of a system.  This is at the core of the following proposition, but is also suggested by the previous one.  
\begin{Proposition}
A collection of no-signaling extended random variables on a minimal signed probability space is contextual if and only if the contextuality index $\delta$  is greater than zero. 
\end{Proposition}
\begin{proof}
If we assume that the random variables are contextual, this means that there is no non-negative joint probability distribution that explains all the correlations for the random variables. But since they are no-signaling, from \cite{de_barros_negative_2016} it follows that there is a negative probability consistent with the correlations. Since, by definition, $\sum_{\omega\in\Omega}p^*(\omega)=1$, and some of the $p^*(\omega)<0$, it follows that $\sum_{\omega\in\Omega}|p^*(\omega)|>1$, and therefore $\delta \neq 0$. Also, from the definition of negative probabilities, it follows that $\delta$ cannot be less than zero, and we have that $\delta>0$. Now, let us assume that $\delta>0$. Since $\delta$ is the lowest possible value for the L1 norm minus one, this implies that there is no non-negative joint, which also implies contextuality. For a more detailed proof using a different definition of negative probabilities, see~\cite{de_barros_negative_2016}.
\end{proof}

Another straightforward consequence of the definition of negative probabilities is that, for each context $C_i$, the extended random variables are equivalent to regular random variables. This equivalency should not come as a surprise since, for each context, we have a complete data table involving all possible experimental outcomes. We also point out that if there exists a context $C_i$ such that $\Omega_i = \Omega$, then $p^*$ is a probability.

Let us now examine some examples. Let $R_1$, $R_2$, and $R_3$ be three extended random variables defined over a negative probability space, and assume that  \mbox{$C_1=(R_1,R_2)$} and $C_2=(R_1,R_3)$ define two different measurement contexts. 
Then, it follows from Definition \ref{def:negative-probabilities} that \mbox{$p^*(R_1=\alpha)=\sum_{\beta_i} p^*(R_1=\alpha|R_2=\beta_i) p^*(R_2=\beta_i)$} and \mbox{$p^*(R_1=\alpha)=\sum_{\beta_i} p^*(R_1=\alpha|R_3=\beta_i) p^*(R_3=\beta_i)$}, where $\alpha$ and $\beta_i$ are the possible values the random variables can take. In other words, the (pseudo) probability distribution of a random variable defined over a negative probability space cannot depend on whether it is co-observed with one or another random variable  \cite{abramsky_logical_2012,SimulatingNon-signaling,oas_exploring_2014}. As remarked above, this property is known in the physics literature as the ``no-signaling condition'' \cite{popescu_quantum_1994}. Alternatively, if experimental observations of a quantity show its probability distributions as independent of other co-observable variables, then it follows that there always exist a negative probability with extended random variables that model the experimental outcomes. In other words, the existence of extended random variables on a negative probability space is a necessary and sufficient condition for the non-signaling condition to hold \cite{abramsky_logical_2012,SimulatingNon-signaling,oas_exploring_2014}.

The equivalence between negative probabilities and non-signaling is one reason why negative probabilities may be a useful tool for exploring the quantum world. Additionally, other properties of quantum systems are well described by negative probabilities. For example, in reference \cite{oas_survey_2015}, many of the principles attempted to describe quantum mechanics were represented in terms of negative probabilities. It was shown there that negative probabilities provided an elegant and straightforward way to express them. 

At this point, it is illustrative to consider the example of two photons, $A$ and $B$, in the singlet state with $z$-polarization either $\pm 1$, given by (\ref{eq:Bell-entangled}).  We saw in Section \ref{sec:contextuality-in-QM} that no probability distribution exists that can account for the quantum correlations, because quantum mechanics violates (\ref{eq:S-inequality}). However, let~us see how we can build a negative probability distribution for the above example. First, we point out that for the above case, the smallest $\Omega$ we can use, without loss of generality \cite{suppes_when_1981}, is given by 
\begin{equation}
\Omega = \{ \omega_{\bar{a}\bar{a}'\bar{b}\bar{b}'},\omega_{ \bar{a}\bar{a}'\bar{b}b'},\omega_{ \bar{a}\bar{a}'b\bar{b}'},\omega_{ \bar{a}\bar{a}'bb'},\omega_{\bar{a}a'\bar{b}\bar{b}'},\ldots,\omega_{aa'b\bar{b}'},\omega_{aa'bb'} \},    
\end{equation}
where $\omega_{aa'bb'}$ corresponds to the outcome $A=a$, $A'=a'$, $B=b$, and $B'=b'$. It should be clear that $\Omega$ generates a $\sigma$-algebra $\mathcal{F}$, formed by all its subsets (i.e., $\mathcal{F}=\mathcal{P}(\Omega)$). Accordingly, the random variables can be defined easily from $\Omega$. For example, $A$ would be the random variable defined as the following~function.
\begin{equation}
    \label{eq:A-random-variable}
    A(\omega) =
    \left\{
	\begin{array}{ll}
		+1  & \mbox{if } \omega \in \{\omega_{a\bar{a}'\bar{b}\bar{b}'}, \omega_{a\bar{a}'\bar{b}b'},\omega_{a\bar{a}'b\bar{b}'},\omega_{a\bar{a}'bb'},\omega_{aa'\bar{b}\bar{b}'},\omega_{aa'\bar{b}b'},\omega_{aa'b\bar{b}'},\omega_{aa'bb'}\} \\
		-1 &  \mbox{if } \omega \in \{\omega_{\bar{a}\bar{a}'\bar{b}\bar{b}'},\omega_{ \bar{a}\bar{a}'\bar{b}b'},\omega_{\bar{a}\bar{a}'b\bar{b}'},\omega_{\bar{a}\bar{a}'bb'},\omega_{ \bar{a}a'\bar{b}\bar{b}'},\omega_{a\bar{a}'b\bar{b}'},\omega_{\bar{a}a'b\bar{b}'},\omega_{\bar{a}a'bb'}\}
	\end{array}.
\right.
\end{equation}
Alternatively, $A'$ is given by
\begin{equation}
    \label{eq:Ap-random-variable}
    A'(\omega) =
    \left\{
	\begin{array}{ll}
		+1  & \mbox{if } \omega \in \{\omega_{\bar{a}a'\bar{b}\bar{b}'},\omega_{\bar{a}a'\bar{b}b'},\omega_{\bar{a}a'b\bar{b}'},\omega_{ \bar{a}a'bb'},\omega_{aa'\bar{b}\bar{b}'},\omega_{aa'\bar{b}b'},\omega_{aa'b\bar{b}'},\omega_{aa'bb'}\} \\
		-1 & \mbox{if } \omega \in \{ \omega_{\bar{a}\bar{a}'\bar{b}\bar{b}'},\omega_{\bar{a}\bar{a}'\bar{b}b'},\omega_{\bar{a}\bar{a}'b\bar{b}'},\omega_{\bar{a}\bar{a}'bb'},\omega_{a\bar{a}'\bar{b}\bar{b}'}, \omega_{a\bar{a}'\bar{b}b'},\omega_{a\bar{a}'b\bar{b}'},\omega_{a\bar{a}'bb'} \}
	\end{array},
\right.
\end{equation}
and similarly for $B$ and $B'$. On the other hand, given that $A$ and $B$ are compatible in the two photons model, there exists a context that contains both. This means that there exists an observable $(A,B)$, that~gives the joint outcomes $(i,j)$ ($i,j=\pm 1$) of performing a simultaneous measure of both $A$ and $B$. It is defined by

\begin{equation}
    \label{eq:(A,B)-random-variable}
    (A,B)(\omega) =
    \left\{
	\begin{array}{llll}
		(+1,+1)  & \mbox{if } \omega \in \{\omega_{aa'bb'},\omega_{a\bar{a}'bb'},\omega_{aa'b\bar{b}'},\omega_{a\bar{a}'b\bar{b}'} \} \\
		(-1,+1) & \mbox{if } \omega \in \{ \omega_{ \bar{a}a'bb'},\omega_{\bar{a}\bar{a}'bb'},\omega_{\bar{a}a'b\bar{b}'},\omega_{\bar{a}\bar{a}'b\bar{b}'}\}\\
		(+1,-1) & \mbox{if } \omega \in \{ \omega_{aa'\bar{b}b'},\omega_{aa'\bar{b}\bar{b}'},\omega_{a\bar{a}'\bar{b}b'},\omega_{a\bar{a}'\bar{b}\bar{b}'}\}\\
		(-1,-1) & \mbox{if } \omega \in \{ \omega_{ \bar{a}a'\bar{b}b'},\omega_{\bar{a}\bar{a}'\bar{b}b'},\omega_{\bar{a}a'\bar{b}\bar{b}'},\omega_{\bar{a}\bar{a}'\bar{b}\bar{b}'}\}
	\end{array}.
\right.
\end{equation}
Let us see how the context defined by $AB$ defines a probability space, and how this space relates to $\Omega$ and $\mathcal{F}$. Notice first that all possible propositions associated to $(A,B)$ (which have the form ``$A$ has value $i$ and $B$ has value $j$'', for $i,j=\pm 1$), are represented by the subsets of $\Omega$ listed in Equation~(\ref{eq:(A,B)-random-variable}). By~computing all possible unions, intersections and complements of these subsets, a~Boolean subalgebra $\mathcal{F}_{(A,B)}$ of $\mathcal{F}$ is generated. Now, in a two photons state, $A$ and $B$ are of course compatible, and there exists a probability assignment (defined by a quantum state of the compound system) $\mu_{(A,B)}$ such that the triad $(\Omega,\mathcal{F}_{(A,B)},\mu_{(A,B)})$ is a classical probability space. If we now consider a global probability assignment $(\Omega,\mathcal{F},\mu)$ (satisfying Definition \ref{def:SignedProba}), if it is a valid extension, we must have that $\mu(F)=\mu_{(A,B)}(F)$, for all $F\in\mathcal{F}_{(A,B)}$.

Another interesting observable is given by the product of outcomes of $A$ and $B$. Let us denote it by $AB$. It is defined by
\begin{equation}
    \label{eq:AB-random-variable}
    AB(\omega) =
    \left\{
	\begin{array}{ll}
		1  & \mbox{if } \omega \in \{\omega_{aa'bb'},\omega_{a\bar{a}'bb'},\omega_{aa'b\bar{b}'},\omega_{ a\bar{a}'b\bar{b}'}, \omega_{\bar{a}a'\bar{b}b'},\omega_{\bar{a}\bar{a}'\bar{b}b'},\omega_{ \bar{a}a'\bar{b}\bar{b}'},\omega_{\bar{a}\bar{a}'\bar{b}\bar{b}'} \} \\
		-1 & \mbox{if } \omega \in\{\omega_{\bar{a}a'bb'},\omega_{\bar{a}\bar{a}'bb'},\omega_{\bar{a}a'b\bar{b}'},\omega_{\bar{a}\bar{a}'b\bar{b}'},\omega_{aa'\bar{b}b'},\omega_{aa'\bar{b}\bar{b}'},\omega_{a\bar{a}'\bar{b}b'},\omega_{a
		\bar{a}'\bar{b}\bar{b}'}\}
	\end{array}.
\right.
\end{equation}
We obtain again a Boolean subalgebra $\mathcal{F}_{AB}$ of $\mathcal{F}$. Similar constructions can be made for $A'B$, $AB'$, $AA'$, $BB'$, $(A,A')$, $(A,B')$, and so on. What are the differences between those observables that mix incompatible observables (such as $AA'$) with respect to those which do not (such as $AB$)? If we write down the details for $AA'$, we obtain
\begin{equation}
    \label{eq:AA'-random-variable}
    AA'(\omega) =
    \left\{
	\begin{array}{ll}
		1  & \mbox{if } \omega \in \{\omega_{aa'bb'},\omega_{aa'b\bar{b}'},\omega_{aa'\bar{b}b'},\omega_{aa'\bar{b}\bar{b}'},\omega_{\bar{a}\bar{a}'\bar{b}b'},\omega_{\bar{a}\bar{a}'\bar{b}\bar{b}'}, \omega_{\bar{a}\bar{a}'bb'},\omega_{\bar{a}\bar{a}'b\bar{b}'}\} \\
		-1 & \mbox{if } \omega \in \{\omega_{\bar{a}a'bb'},\omega_{\bar{a}a'b\bar{b}'},\omega_{\bar{a}a'\bar{b}b'},\omega_{\bar{a}a'\bar{b}\bar{b}'},\omega_{a\bar{a'}\bar{b}b'},\omega_{a\bar{a'}\bar{b}\bar{b}'},\omega_{a\bar{a'}bb'},\omega_{a\bar{a'}b\bar{b}'}\}
	\end{array}.
\right.
\end{equation}
We get again a Boolean subalgebra $\mathcal{F}_{AA'}$ for $AA'$. Notice first that $\mathcal{F}_{AA'}\neq\mathcal{F}_{AB}$. Second, if we want to define probabilities for the outcomes of $AA'$, we have to consider the measures defined by the model we are considering, here a two photons system. In this case, the states are determined by the Born rule. We know that if a collection of observables is commutative, a quantum state assigns them a positive probability. Thus, any legitimate quantum state will assign positive probabilities for all the events in the Boolean algebras $\mathcal{F}_{AB}$, $\mathcal{F}_{AB'}$, $\mathcal{F}_{A'B}$ and $\mathcal{F}_{A'B'}$. What happens with the events in $\mathcal{F}_{AA'}$ and $\mathcal{F}_{BB'}$? The non-negativity of the probabilities assigned by quantum states to the propositions associated with those algebras is no longer granted. This will become clear with the examples discussed in the following Section (see Proposition \ref{prop:four-variables}).

Quantum mechanics tells us that, in addition to the correlations in (\ref{eq:Tsirelson-bound}), the observable expectations also satisfy the following:
\begin{equation}
\label{eq:Bell-EPR-marginals}
    \langle A \rangle =\langle A' \rangle =\langle B \rangle =\langle B' \rangle = 0.
\end{equation}
If we now impose (\ref{eq:Tsirelson-bound}) and (\ref{eq:Bell-EPR-marginals}) to the probabilities, from the definition of the random variables set above, we would have at once that the probabilities of $\omega_i$ would have to satisfy the following set of linear equations.

\begin{equation}
\label{eq:prob-sum-1}
\begin{array}{ccc}
     p_{\bar{a}\bar{a}'\bar{b}\bar{b}'}+p_{\bar{a}\bar{a}'\bar{b}b'}+p_{\bar{a}\bar{a}'b\bar{b}'}+p_{\bar{a}\bar{a}'bb'}+p_{\bar{a}a'\bar{b}\bar{b}'}+p_{\bar{a}a'\bar{b}b'}+p_{\bar{a}a'b\bar{b}'}+p_{\bar{a}a'bb'} & &\\
     +p_{a\bar{a}'\bar{b}\bar{b}'}+p_{a\bar{a}'\bar{b}b'}+p_{a\bar{a}'b\bar{b}'}+p_{a\bar{a}'bb'}+p_{aa'\bar{b}\bar{b}'}+p_{aa'\bar{b}b'}+p_{aa'b\bar{b}'}+p_{aa'bb'} & = & 1,
     \end{array}
\end{equation}
\begin{equation}
\label{eq:expAzero}
\begin{array}{ccc}
-p_{\bar{a}\bar{a}'\bar{b}\bar{b}'}-p_{\bar{a}\bar{a}'\bar{b}b'}-p_{\bar{a}\bar{a}'b\bar{b}'}-p_{\bar{a}\bar{a}'bb'}-p_{\bar{a}a'\bar{b}\bar{b}'}-p_{\bar{a}a'\bar{b}b'}-p_{\bar{a}a'b\bar{b}'}-p_{\bar{a}a'bb'} & &\\
     +p_{a\bar{a}'\bar{b}\bar{b}'}+p_{a\bar{a}'\bar{b}b'}+p_{a\bar{a}'b\bar{b}'}+p_{a\bar{a}'bb'}+p_{aa'\bar{b}\bar{b}'}+p_{aa'\bar{b}b'}+p_{aa'b\bar{b}'}+p_{aa'bb'} & = & 0,
\end{array}
\end{equation}
\begin{equation}
\label{eq:expApzero}
\begin{array}{ccc}
     -p_{\bar{a}\bar{a}'\bar{b}\bar{b}'}-p_{\bar{a}\bar{a}'\bar{b}b'}-p_{\bar{a}\bar{a}'b\bar{b}'}-p_{\bar{a}\bar{a}'bb'}+p_{\bar{a}a'\bar{b}\bar{b}'}+p_{\bar{a}a'\bar{b}b'}+p_{\bar{a}a'b\bar{b}'}+p_{\bar{a}a'bb'} & &\\
     -p_{a\bar{a}'\bar{b}\bar{b}'}-p_{a\bar{a}'\bar{b}b'}-p_{a\bar{a}'b\bar{b}'}-p_{a\bar{a}'bb'}+p_{aa'\bar{b}\bar{b}'}+p_{aa'\bar{b}b'}+p_{aa'b\bar{b}'}+p_{aa'bb'} & = & 0,
\end{array}
\end{equation}
\begin{equation}
\label{eq:expBzero}
\begin{array}{ccc}
     -p_{\bar{a}\bar{a}'\bar{b}\bar{b}'}-p_{\bar{a}\bar{a}'\bar{b}b'}+p_{\bar{a}\bar{a}'b\bar{b}'}+p_{\bar{a}\bar{a}'bb'}-p_{\bar{a}a'\bar{b}\bar{b}'}-p_{\bar{a}a'\bar{b}b'}+p_{\bar{a}a'b\bar{b}'}+p_{\bar{a}a'bb'} & &\\
     -p_{a\bar{a}'\bar{b}\bar{b}'}-p_{a\bar{a}'\bar{b}b'}+p_{a\bar{a}'b\bar{b}'}+p_{a\bar{a}'bb'}-p_{aa'\bar{b}\bar{b}'}-p_{aa'\bar{b}b'}+p_{aa'b\bar{b}'}+p_{aa'bb'} & = & 0,
\end{array}
\end{equation}
\begin{equation}
\label{eq:expBpzero}
\begin{array}{ccc}
-p_{\bar{a}\bar{a}'\bar{b}\bar{b}'}+p_{\bar{a}\bar{a}'\bar{b}b'}-p_{\bar{a}\bar{a}'b\bar{b}'}+p_{\bar{a}\bar{a}'bb'}-p_{\bar{a}a'\bar{b}\bar{b}'}+p_{\bar{a}a'\bar{b}b'}-p_{\bar{a}a'b\bar{b}'}+p_{\bar{a}a'bb'} & &\\
     -p_{a\bar{a}'\bar{b}\bar{b}'}+p_{a\bar{a}'\bar{b}b'}-p_{a\bar{a}'b\bar{b}'}+p_{a\bar{a}'bb'}-p_{aa'\bar{b}\bar{b}'}+p_{aa'\bar{b}b'}-p_{aa'b\bar{b}'}+p_{aa'bb'} & = & 0,
\end{array}
\end{equation}
\begin{equation}
\label{eq:expAB}
\begin{array}{ccc}
p_{\bar{a}\bar{a}'\bar{b}\bar{b}'}+p_{\bar{a}\bar{a}'\bar{b}b'}-p_{\bar{a}\bar{a}'b\bar{b}'}-p_{\bar{a}\bar{a}'bb'}+p_{\bar{a}a'\bar{b}\bar{b}'}+p_{\bar{a}a'\bar{b}b'}-p_{\bar{a}a'b\bar{b}'}-p_{\bar{a}a'bb'} & &\\
     -p_{a\bar{a}'\bar{b}\bar{b}'}-p_{a\bar{a}'\bar{b}b'}+p_{a\bar{a}'b\bar{b}'}+p_{a\bar{a}'bb'}-p_{aa'\bar{b}\bar{b}'}-p_{aa'\bar{b}b'}+p_{aa'b\bar{b}'}+p_{aa'bb'} & = & \frac{1}{\sqrt{2}},
\end{array}
\end{equation}
\label{eq:expABp}
\begin{equation}
\begin{array}{ccc}
p_{\bar{a}\bar{a}'\bar{b}\bar{b}'}-p_{\bar{a}\bar{a}'\bar{b}b'}+p_{\bar{a}\bar{a}'b\bar{b}'}-p_{\bar{a}\bar{a}'bb'}+p_{\bar{a}a'\bar{b}\bar{b}'}-p_{\bar{a}a'\bar{b}b'}+p_{\bar{a}a'b\bar{b}'}-p_{\bar{a}a'bb'} & &\\
     -p_{a\bar{a}'\bar{b}\bar{b}'}+p_{a\bar{a}'\bar{b}b'}-p_{a\bar{a}'b\bar{b}'}+p_{a\bar{a}'bb'}-p_{aa'\bar{b}\bar{b}'}+p_{aa'\bar{b}b'}-p_{aa'b\bar{b}'}+p_{aa'bb'} & = & \frac{1}{\sqrt{2}},
\end{array}
\end{equation}
\begin{equation}
\label{eq:expApB}
\begin{array}{ccc}
p_{\bar{a}\bar{a}'\bar{b}\bar{b}'}+p_{\bar{a}\bar{a}'\bar{b}b'}-p_{\bar{a}\bar{a}'b\bar{b}'}-p_{\bar{a}\bar{a}'bb'}-p_{\bar{a}a'\bar{b}\bar{b}'}-p_{\bar{a}a'\bar{b}b'}+p_{\bar{a}a'b\bar{b}'}+p_{\bar{a}a'bb'} & &\\
     +p_{a\bar{a}'\bar{b}\bar{b}'}+p_{a\bar{a}'\bar{b}b'}-p_{a\bar{a}'b\bar{b}'}-p_{a\bar{a}'bb'}-p_{aa'\bar{b}\bar{b}'}-p_{aa'\bar{b}b'}+p_{aa'b\bar{b}'}+p_{aa'bb'} & = & \frac{1}{\sqrt{2}},
\end{array}
\end{equation}
\begin{equation}
\label{eq:expApBp}
\begin{array}{ccc}
p_{\bar{a}\bar{a}'\bar{b}\bar{b}'}-p_{\bar{a}\bar{a}'\bar{b}b'}+p_{\bar{a}\bar{a}'b\bar{b}'}-p_{\bar{a}\bar{a}'bb'}-p_{\bar{a}a'\bar{b}\bar{b}'}+p_{\bar{a}a'\bar{b}b'}-p_{\bar{a}a'b\bar{b}'}+p_{\bar{a}a'bb'} & &\\
     +p_{a\bar{a}'\bar{b}\bar{b}'}-p_{a\bar{a}'\bar{b}b'}+p_{a\bar{a}'b\bar{b}'}-p_{a\bar{a}'bb'}-p_{aa'\bar{b}\bar{b}'}+p_{aa'\bar{b}b'}-p_{aa'b\bar{b}'}+p_{aa'bb'} & = & -\frac{1}{\sqrt{2}},
\end{array}
\end{equation}
where we are using the simplifying notation that $p_{aa'bb'}=p^*(\omega_{aa'bb'})$, $p_{aa'b\bar{b}'}=p^*(\omega_{aa'b\bar{b}'})$, and so on. Notice that Equation (\ref{eq:prob-sum-1}) corresponds to the condition $\mu(\Omega)=1$ in Definition \ref{def:SignedProba}. \mbox{Equations (\ref{eq:expAzero})--(\ref{eq:expBpzero})} represent the expectations in (\ref{eq:Bell-EPR-marginals}). Finally, Equations (\ref{eq:expAB})--(\ref{eq:expApBp}) are the expectations computed using~(\ref{eq:Tsirelson-bound}).

Equations (\ref{eq:prob-sum-1})--(\ref{eq:expApBp}) form a set of nine linearly independent equations. However, to completely determine the probabilities of each the 16 elementary events $\omega_i\in\Omega$, one needs a total of 16 equations. Thus, the problem is under-determined. However, it is possible to write a general solution to (\ref{eq:prob-sum-1})--(\ref{eq:expApBp}) that will have seven undetermined parameters, and it is straightforward to show that at least one of the $p_{\omega_i}$'s are negative for all possible solutions. But if one compute the marginal expectations for each of the experimental contexts, one would observe that for contexts $C_1 = (A,B)$, $C_2 = (A,B')$, $C_3 = (A',B)$, and $C_4 = (A',B)$ all the marginal probabilities are non-negative. What we mean is that the marginal probabilities observed in, say, $C_1$, i.e. $p^*(A=\pm 1,B=\pm 1)$, are all non-negative. This~comes from the constraints in (\ref{eq:prob-sum-1})--(\ref{eq:expApBp}). An explicit solution to (\ref{eq:prob-sum-1})--(\ref{eq:expApBp}) is lengthy and cumbersome but can be obtained easily. The interested reader can either examine a solution given in reference \cite{de_barros_negative_2016} or compute it themselves. 

We now prove a general relationship between quantum mechanics and negative probabilities. 
\begin{Proposition}
\label{prop:quantum-is-negative}
Let $\mathcal{Q}$ be the collection of complete sets of simultaneously observable one-dimensional projection operators on a Hilbert space $\mathcal{H}$, i.e., for each $Q_i\in\mathcal{Q}$  there are $N=\dim \mathcal{H}$ commuting projection operators such that $\sum_{\hat{P}_j\in Q_i}\hat{P}_j=\hat{1}$. Let $p$ be a measure over elements of $Q_i$ given by Born's rule. Let also $\{ R^{*}_i \}$ be a collection of extended dichotomous random variables on a signed measure space $(\Omega,\mathcal{F},\mu)$, such that for each $Q_i$  there is a context $C_i$  such that for all $\hat{P}_j\in Q_i$ there is a 1-1 equivalent element of $C_i$ with the same marginal probability distributions, i.e., within a context $C_i$ the expectations of $R^*_j$ and $\hat{P}_j$ are the same, as well as any other higher moments in combination with other variables in the same context. Then $\mu$ is a negative probability space that represents all contexts $C_i$.
\end{Proposition}
\begin{proof}
To prove that $\mu$ is negative probability space, we just need to show that  $\mu(\Omega)=1$. In order do so, let us notice that each extended random variable $R^*_i$ defines a partition of the sample space $\Omega$ corresponding to each of their values (similarly to what we had in Equations  (\ref{eq:A-random-variable})--(\ref{eq:AA'-random-variable})). For each combination of extended random variables, there is a corresponding partition. In particular, for a given projection operator, say, $\hat{P}_1$, by assumption, there exists a two-valued extended random variable $R^*_1$. The two outcomes, $R^*_1=1$ and $R^*_1=-1$, define a partition of $\Omega$, formed by two subsets that we denote by $F_{1}$ and $F_{-1}$, such that $F_{1}\cap F_{-1}=\emptyset$ and $F_{1}\cup F_{-1}=\Omega$. Since the measure $\mu$ assigns to those subsets the same probabilities as the Born's rule, we must have $1= \langle\hat{P}_1\rangle+\langle\hat{1}-\hat{P}_1\rangle =\mu(F_{1})+ \mu(F_{-1})=\mu(\Omega)$. Thus, $\mu$ is normalized, and defines a negative probability.
\end{proof}

In the following section we present, in Propositions \ref{prop:three-variables} and \ref{prop:four-variables}, examples of how this result applies in simple but important cases. We end with this section with a final comment. The requirement that $p^*$ minimizes the L1 norm (see Definition \ref{def:negative-probabilities}) provides us with a number $\delta$ that is greater than or equal to zero. If it is zero, the random variables are not contextual, and proper a joint probability distribution 
exists. However, the correlations for the Bell-EPR case do not allow for a proper joint 
\cite{fine_joint_1982}. The fact that $\delta$ is not zero provides a way for measuring how contextual (or, in this 
case, because it is contextual-at-a-distance, how non-local) a system of random variables is. The more $\delta$ 
departs from $0$, the more contextual it is 
\cite{de_barros_unifying_2014,de_barros_measuring_2015,dzhafarov_generalizing_2014,dzhafarov_contextuality-by-default_2017}.

In this section, we showed a generalized probability theory that includes negative (or signed) probabilities. This theory is well suited for describing quantum systems, as it is compatible with the no-signaling condition. Furthermore, negative probabilities have advantages with other alternative extended probability theories. For example, upper and lower probabilities can also be used to describe quantum contextuality \cite{suppes_existence_1991,de_barros_probabilistic_2010}. However, because upper and lower probabilities involve inequalities, their computation is challenging and cumbersome. Additionally, the main appeal for upper and lower probabilities is that they have an interpretation. For instance, monotonic upper and lower probabilities can be interpreted within Dempster-Shaffer theory (they call them plausibility and belief, respectively)~\cite{dempster_upper_1967}. However, this interpretation fails in quantum theory, where upper and lower probabilities are non-monotonic, and Dempster-Shaffer's reasoning does not apply anymore. 

Unlike upper probabilities, negative probabilities can be easily computed, as shown in the example above. Furthermore, one can use negative probabilities as a contextual calculus for conflicting subjective contextual information  even outside of physics \cite{de_barros_decision_2014,de_barros_beyond_2015,de_barros_negative_2014,de_barros_quantum_2015}. So, the use of negative probabilities for quantum systems seem worth exploring.  

However, a question often asked is this: what is the meaning of an event having a negative probability? First, we point out that, in our definition, negative probability events are never observed: negative probabilities exist for the unobserved joint events. This is similar to the use of negative numbers to count physical objects, e.g., apples in a fruit stand. Of course, the concept of a negative number of apples is absurd: one could never observe $-3$ apples. This is emphasized by DeMorgan's comment about negative numbers \cite{morgan_study_1910}: ``[the student] must reject the definition still sometimes given of the quantity $-a$, that it is less than nothing. It is astonishing that the human intellect should ever have tolerated such an absurdity as the idea of a quantity less than nothing; above all, that the notion should have outlived the belief in judicial astrology and the existence of witches, either of which is ten thousand times more possible.'' Even though the meaning may be problematic for DeMorgan, the~use of negative numbers to track operations of future sales and purchases of apples does not need to be; a negative number of apples makes sense, but only as an accounting trick that helps us figure out the observable (non-negative) final number of apples. We do not \emph{need} an interpretation of negative numbers of apples. In this sense, an interpretation of negative probabilities is as unnecessary as an interpretation of negative numbers of apples.  

Nevertheless, there are many different interpretations of negative probabilities for non-monotonic systems (see \cite{abramsky_operational_2014,burgin_interpretations_2010,burgin_introduction_2016,machado_fractional_2013,khrennikov_interpretations_2009,zurek_environment-induced_1982}). For example, Khrennikov  proposes that negative probabilities are associated with sequences that violate von Mises's principle of stability, which states that probabilities are about well-behaved sequences whose mean converge to a certain number \cite{khrennikov_interpretations_2009}. By focusing on infinite sequences that do not converge using the standard real-number metric, Khrennikov showed that such sequences converge using $p$-adic numbers, with negative probabilities being associated to such sequences that violated the principle of stability. Another approach is that of Abramsky and Brandenburger \cite{abramsky_operational_2014}. They proposed to use negative probabilities to describe a data table where events could themselves be signed. In their interpretation, the joint event of, say, three random variables being +1, would also carry an additional bit, a sign. Two events could then cancel each other if their signs were different, and negative probabilities manifest those two types of events.  As mentioned in the previous paragraph, another way to think about negative probabilities is the pragmatic view: negative probabilities are a useful tool for computing quantum probabilities. This view does not demand an interpretation, and it was the way that both Feynman and Dirac thought about negative probabilities~\cite{dirac_bakerian_1942,Feynman-1987}. In this paper, we are proposing that, at least in quantum physics, negative probabilities can be interpreted as a miscounting and mislabeling of a data table because quantum particles, and some propositions about them, are indistinguishable.

\section{Indistinguishability in Quantum Mechanics and Mathematics}
\label{sec:indistinguishability}

Compound quantum systems can be prepared in entangled states that violate non-contextuality inequalities. An example we saw was the state in (\ref{eq:Bell-entangled}), whose correlations (\ref{eq:Bell-EPR-marginals}) lead to a violation of (\ref{eq:S-inequality}). However, there is a different physical effect associated with compound quantum systems involving particles of \emph{the same kind}. To write the state of the compound system, we must invoke the symmetrization postulate. 
This postulate asserts that the state of a compound quantum system of
identical particles must be symmetric under permutation of the
particles if the particles are Bosons and anti-symmetric if they are
Fermions. 

Suppose that we have two Fermions, one of them prepared in the state $|a\rangle$ and the other in the state $|b\rangle$. Then, after applying the symmetrization postulate, the state of the compound system is given~by
\begin{equation}
|\psi\rangle=\frac{1}{\sqrt{2}}\left( |a\rangle\otimes|b\rangle-|b\rangle\otimes|a\rangle\right) .
\end{equation}
A similar procedure should be used to construct the state of two Bosons by using a plus instead of a minus sign, thus yielding a symmetric state.

The implications of the symmetrization postulate (SP) are of significant importance for quantum theory. Pauli's exclusion principle and also the so-called quantum statistics (Einstein-Bose and Fermi-Dirac statistics) follow from the SP. This feature of the quantum formalism is particularly relevant for the study of the properties of indistinguishable particles in quantum information theory~\cite{LoFranco-2018A,Bose-2002,Bose-2013}. Furthermore, the peculiar properties of compound systems of identical particles lead to heated debates in the literature about the interpretation of quantum mechanics. A remarkable position was that of E. Schr\"{o}dinger, who claimed that elementary particles are not individuals, given that the theory gives no means to identify them \cite{schrodinger1950elementary,schroedinger_science_1952}. An even more extraordinary view was that suggested by Wheeler, who~once told Feynman that all electrons have the same properties because they are all the same electron \cite{feynman1963development}. We do not necessarily agree with Schr\"{o}dinger or Wheeler, but we emphasize a broad agreement among physicists that two electrons are indistinguishable at some fundamental level.

Researchers discussed the indistinguishability of elementary particles in connection to indistinguishability in logic and mathematics. Indeed, to deal with genuinely indistinguishable entities, the quasi-set theory was developed as a set-theoretical framework in which the classical laws of identity do not apply for specific elements of the theory (see, for example, \cite{krause_quasi-set_1992,krause_quasi-set_1999,domenech_discussion_2007}). This formalism was used in \cite{domenech_q-spaces_2008,Holik2010} to reconstruct the Fock-space formulation of quantum mechanics avoiding any particle labeling (see \cite{Compagno-2018} for an alternative approach). The axioms of quasi-set theory are chosen so that it is possible to form collections of indistinguishable entities, violating Leibniz's principle of identity of indiscernibles \cite{krause_quasi-set_1992}. In this theory, the identity symbol ``$=$'' cannot be applied to all its elements. Instead, a weaker equivalence relation ``$\equiv$'' is used to describe a situation where an element $x$ is indistinguishable from another element $y$, and it is formally represented by $x\equiv y$. This corresponds to the idea that $x$ and $y$ represent indistinguishable quantum objects. 

Quasi-set theory assumes that a cardinal can be assigned to these collections so that every quasi-set has a definite number of elements. The indistinguishable elements of a quasi-set cannot be identified by names, counted, or ordered. In this sense, the standard set-theory rules do not apply for all elements of the theory. Quasi-sets having indistinguishable elements are thought of as representing collections of quantum objects of the same kind, i.e., indistinguishable objects. Another essential feature of quasi-set theory is that it contains a copy of Zermelo-Fraenkel set theory to develop standard mathematics within it.  

Quasi-set theory allows us to formally describe collections of indiscernible objects without resorting to any mathematical tricks. The connection between indistinguishability and contextuality was studied recently. In \cite{de_barros_contextuality_2017}, we have shown that the possibility of identifying particles in different contexts lies at the core of the Kochen-Specker contradiction. In \cite{de_barros_indistinguishability_2019}, we studied how the assumption of the indistinguishability of properties allows one to understand the occurrence of contextual random~variables. 

The connection between particle indistinguishability and indistinguishability of properties is essential here. So, let us examine how it comes about. In the quantum formalism, a testable proposition about an object is formally represented by a projection operator. Given an observable $A$, consider the proposition ``the value of $A$ lies in the interval $\Delta$'' (that we write compactly as $P^{A}(\Delta)$). By using the spectral theorem, $P^{A}(\Delta)$ can be mathematically represented by an orthogonal projection $\hat{P}^{A}(\Delta)$ (notice that the ``hat'' distinguishes the mathematical object from the proposition it represents). We aim to represent quantum properties related to the particles and describe expressions such as ``a particle has a certain property.''

It is instructive to illustrate the connection between quantum indistinguishability and the identification of propositions with the same content, but in different contexts, by considering a quasi-pair concept in quasi-set theory.  The quasi-pair  $\langle [x],P^{A}(\Delta)\rangle$ can be used to describe one quanta possessing the property $P^{A}(\Delta)$ (see also the discussion presented in \cite{de_barros_indistinguishability_2019}), where the $[x]$ is the collection of all possible indistinguishable elements from $x$. Thus, $\langle [x],P^{A}(\Delta)\rangle$, can be interpreted as: ``a quantum object satisfies that the value of $A$ lies in $\Delta$''. Notice that we refer to a quantum object, without specifying which one it is (because, according to the spirit of quasi-set theory, they are indiscernible). The classical analog of this proposition could make explicit reference to the particle identity (as, for example, in ``particle $e_{1}$ satisfies that the value of $A$ lies in $\Delta$''). Moreover, we could use standard set theory and write $\langle \{e_{1}\},P^{A}(\Delta)\rangle$ (notice that, in the last pair, we are using the standard singleton $\{e_{1}\}$, which is formed by the sole individual $e_{1}$). However, this is impossible if we assume that quantum particles are indistinguishable, and we use quasi-set theory. If we now take another quanta $y$ such that $y\equiv x$, and consider the proposition $\langle [y],P^{A}(\Delta)\rangle$, using the rules of quasi-set theory, we obtain $\langle [x],P^{A}(\Delta)\rangle\equiv\langle [y],P^{A}(\Delta)\rangle$. This can be interpreted as follows: \textit{indistinguishability of particles leads to the identification of propositions among different contexts}. Each time we consider different instances of a proposition about a quantum system, the propositions associated with these instances are indistinguishable, and thus, they can be identified.  Notice that a proportion's instantiation has the form ``a quantum object's value of $A$ lies in $\Delta$.'' If we now have an instantiation of an equivalent assertion, but considered in a different context, given that we cannot refer to the identity of the quanta involved, we have no means to distinguish the propositions either. Assuming the axioms given in \cite{krause_quasi-set_1992}, indistinguishable quasi-sets are identical (but have in mind that, in this framework, identity is a \textit{derived notion}). It is in this sense that indistinguishable propositions can be identified. 

The above discussion is particularly relevant for the problem mentioned at the end of Section~\ref{sec:contextuality-in-QM}. Given the random variables $A_{B}$ and $A_{B'}$ discussed in Section \ref{sec:contextuality-in-QM} (that have the same content), we~have two options: either $A_{B}=A_{B'}$, or $A_{B}\neq A_{B'}$. Assuming that quanta are indistinguishable and describing propositions using quasi-set theory (as above), when all propositions associated to $A_{B}$ have indistinguishable counterparts in those associated to $A_{B'}$, we obtain that $A_{B}\equiv A_{B'}$ (i.e., they can be identified as random variables). The assumption of quanta indistinguishability, together with the use of quasi-set theory, serves as a justification for identifying those random variables (see \cite{de_barros_indistinguishability_2019,de_barros_contextuality_2017} for a related discussion). 

Let us now use the above framework to connect particle indistinguishability with non-signaling. Let $\mathbf{A}$ and $\mathbf{B}$ represent two agents, Alice and Bob, that aim to communicate with each other. For $\mathbf{A}$  to send a signal to $\mathbf{B}$, they need to appeal to some physical mechanism that can be generally described by sharing a physical system that induces observable correlations between what they observe on it. Suppose that they can measure different observables on their respective sides. We denote by $A$, $A'$, etc., the observables for $\textbf{A}$, and $B$, $B'$, etc. for $\textbf{B}$). Given $A$ and $A'$, we assume that they are complementary, i.e., that if Alice selects $A$, she cannot at the same time select $A'$; similarly for Bob's $B$ and $B'$. However, because Alice and Bob are observing different parts of the communication device, we assume that any of the observables for $\textbf{A}$ are always compatible with whatever choice Bob makes in $\textbf{B}$. The idea of a communication device is that Alice can affect Bob's observations of $B$ or $B'$ by changing her settings from observing $A$ to $A'$ (or vice versa), . 

Let us assume now that Alice and Bob construct a device that works. In other words, they figured out a way to communicate between themselves using some (unknown to us) mechanism where Alice's choices affect Bob's observations. However, Alice and Bob now make a new proposal: they want to see if their device works with indistinguishable quantum particles. This proposal means that whenever we have the contexts $(A, B)$ and $(A, B')$, the properties associated with $A$ in context $B$ are indistinguishable from those of $A$ in context $B'$. Under these assumptions, we should have that, for each property, the~probability of obtaining $P^{A}(\Delta)$ in context $B$ is the same as the probability of obtaining $P^{A}(\Delta)$ in context $B'$. If they were not the same, Alice could use these probabilities to attach an ``identity card'' to some particles in $B$ but not to others. This would be a way of distinguishing indistinguishable particles. 

The above conclusion leads to the following conditions:
\begin{equation}
\label{eq:no-signal-1}
    \sum_{b}p(P^{A}(a),P^{B}(b)|A,B)= \sum_{b}p(P^{A}(a),P^{B'}(b)|A,B')=p(P^{A}(a)|A)
\end{equation}
and 
\begin{equation}
\label{eq:no-signal-2}
    \sum_{a}p(P^{A}(a),P^{B}(b)|A,B)= \sum_{a}p(P^{A'}(a),P^{B}(b)|A',B)=p(P^{B}(b)|B).
\end{equation}
Equations (\ref{eq:no-signal-1}) and (\ref{eq:no-signal-2}) are no-signaling conditions \cite{popescu_quantum_1994}. Thus, the assumption of indistinguishability of properties leads to the no-signaling condition: whatever Alice does to ``her particle'' cannot affect what Bob infers about ``his particle,'' because this would mean attaching an identity card to Alice's and Bob's particles. This condition is extreme, and is specific to physical theories, in particular quantum mechanics, and should not hold in other domains (such as cognition; see, for example \cite{de_barros_quantum_2015,de_barros_beyond_2015,cervantes_snow_2018}). 

To summarize, quantum particles are indistinguishable, and this indistinguishability leads to the indistinguishability of properties. However, we showed that property indistinguishability implies that communication devices such as those discussed by \cite{dieks_communication_1982} cannot work. If we could use the correlations in entangled systems to send a signal between Alice and Bob, such devices could distinguish particles. 

Let us consider two examples that illustrate how the following chain of implications works.

\begin{equation*}
    \mbox{Indistinguishability}\Longrightarrow\mbox{No-signal}\Longrightarrow\mbox{Negative probabilities}
\end{equation*}
We illustrate the above idea with Propositions \ref{prop:three-variables} and \ref{prop:four-variables}. Below we go through the proof of \mbox{Propositions \ref{prop:three-variables} and  \ref{prop:four-variables}}, but we stress that the proofs are all based on the idea put forth above, namely that indistinguishability implies no-signaling, and therefore negative probabilities.  Let us first clarify the notation. Consider three dichotomous random variables forming jointly measurable pairs $X-Y$, $X-Z$, and $Y-Z$. We denote by $X_Y$ the random variable $X$ in the context $X-Y$, with a similar interpretation for $X_Z$, $Y_X$, $Y_Z$, $Z_X$, and $Z_Y$.  Then, we have the following proposition, whose proofs follow the above idea that indistinguishability implies no-signaling, which implies negative probabilities.

%\begin{Proposition}
%\label{prop:three-variables}
%Let $X$, $X-Z$, and $Y-Z$ be three contexts for which $X_Y$, $X_Z$, $Y_X$, $Y_Z$, $Z_X$, and $Z_Y$ are dichotomous random variables. If $X_Y\equiv X_Z$, $Y_X\equiv Y_Z$, and $Z_X\equiv Z_Y$, then, there exists a signed probability space satisfying Definition \ref{def:negative-probabilities}, for which each pair of jointly measurable variables is a context.
%\end{Proposition}

%\begin{Proposition}
%Let $X-Y$, $X-Z$, and $Y-Z$ be three contexts for which $X_Y$, $X_Z$, $Y_X$, $Y_Z$, $Z_X$, and $Z_Y$ are dichotomous random variables. If $X_Y\equiv X_Z$, $Y_X\equiv Y_Z$, and $Z_X\equiv Z_Y$, then, there exists a signed probability space satisfying Definition \ref{def:negative-probabilities}, for which each pair of jointly measurable variables is a context.
%\end{Proposition}

\begin{Proposition}
\label{prop:three-variables}
For jointly measurable pairs $X-Y$, $X-Z$ and $Y-Z$ of dichotomous random variables, if the indistinguishability relations $X_Y\equiv X_Z$, $Y_X\equiv Y_Z$, and $Z_X\equiv Z_Y$ are satisfied, there exists a signed probability space (i.e., satisfying Definition \ref{def:SignedProba}), for which each pair of jointly measurable variables is a context (satisfying Definition \ref{def:NEWcontext}).
\end{Proposition}

\begin{proof}
Suppose that we have three dichotomous random variables, $X$, $Y$ and $Z$. Assume that $X-Y$, $X-Z$ and $Y-Z$, are jointly measurable. In the context $X-Y$, we have different elementary events, which are given by $X=1$ and $Y=1$, $X=-1$ and $Y=1$, $X=1$ and $Y=-1$, and $X=-1$ and $Y=-1$. Let us denote these results by $xy$, $x\bar{y}$, $\bar{x}y$ and $\bar{x}\bar{y}$, respectively. Combined propositions are given by sets like $\{xy,x\bar{y}\}$ (representing the proposition ``$xy$ or $x\bar{y}$''), $\{x\bar{y},\bar{x}y,\bar{x}\bar{y}\}$ (representing ``not $xy$''), and so on. If we define $X-Y:=\{xy,x\bar{y},\bar{x}y,\bar{x}\bar{y}\}$, the complete Boolean algebra is given by $\mathcal{P}(X-Y)$ (that we denote by $\mathcal{B}_{X;Y}$) and can be represented by the diagram in Figure \ref{fig:Hasse}.
\begin{figure}[ht]
\begin{center}
\begin{tikzpicture}
  \node (max) at (0,6) {$\{xy,x\bar{y},\bar{x}y,\bar{x}\bar{y}\}$};
  \node (u1) at (-5,2) {$\{xy,x\bar{y}\}$};
  \node (u2) at (-3,2) {$\{xy,\bar{x}y\}$};
  \node (u3) at (-1,2) {$\{xy,\bar{x}\bar{y}\}$};
  \node (u4) at (1,2) {$\{x\bar{y},\bar{x}y\}$};
  \node (u5) at (3,2) {$\{x\bar{y},\bar{x}\bar{y}\}$};
  \node (u6) at (5,2) {$\{\bar{x}y,\bar{x}\bar{y}\}$};
  \node (uu1) at (-4,4) {$\{xy,\bar{x}y,x\bar{y}\}$} ;
  \node (uu2) at (-1.5,4) {$\{xy,\bar{x}y,\bar{x}\bar{y}\}$};
  \node (uu3) at (1.5,4) {$\{xy,x\bar{y},\bar{x}\bar{y}\}$};
  \node (uu4) at (4,4) {$\{x\bar{y},\bar{x}y,\bar{x}\bar{y}\}$};
  \node (g1) at (-2,0) {$\{xy\}$};
  \node (g2) at (-0.7,0) {$\{x\bar{y}\}$};
  \node (g3) at (0.7,0) {$\{\bar{x}y\}$};
  \node (g4) at (2,0) {$\{\bar{x}\bar{y}\}$};
  \node (min) at (0,-2) {$\emptyset$};
  \draw  (g4) -- (min) -- (g1) -- (u1) -- (uu3) -- (max)
  
(max) -- (uu1) -- (u2) -- (g3) -- (u6)
  
(max) -- (uu3) -- (u1) -- (g2) -- (min) -- (g3)  -- (u4) -- (uu1)
   
(u1) -- (uu1)
  
(u6) -- (uu4) -- (max)

(uu2) -- (u6) -- (g4) -- (u5) -- (uu4)

(g4) -- (u3) -- (uu3)

(u3) -- (uu2) -- (max) 

(u3) -- (g1) -- (u2) -- (uu2)

(uu3) -- (u5) -- (g2) -- (u4) -- (uu4);

  %\draw[preaction={draw=white, -,line width=6pt}] (a) -- (e) -- (c);
\end{tikzpicture}
\end{center}
\caption{Hasse diagram of the $X-Y$ Boolean algebra.\label{fig:Hasse}}
\end{figure}
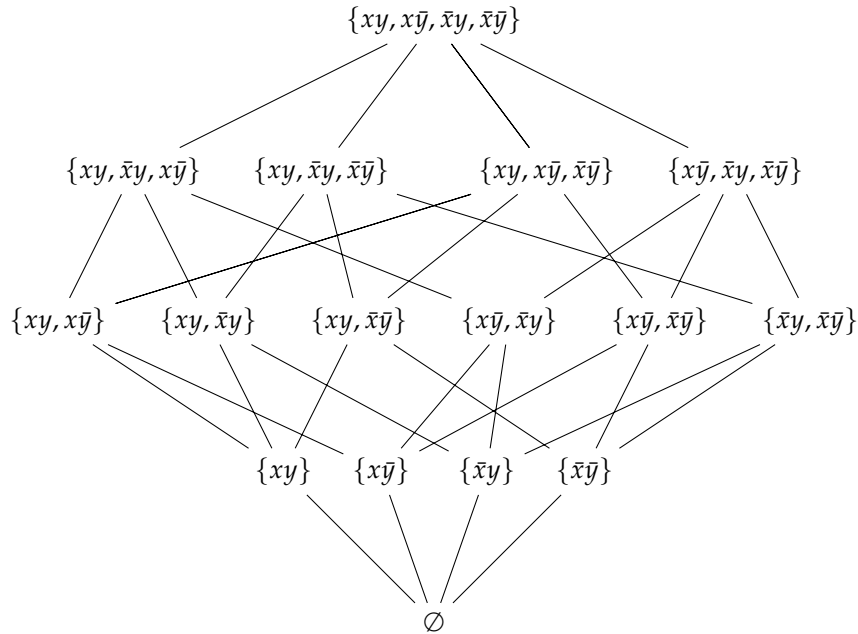

Analogous Boolean algebras $\mathcal{B}_{X;Z}$ and $\mathcal{B}_{Y;Z}$ hold for $X-Z$ and $Y-Z$, which are given by all possible subsets of $\{xz,x\bar{z},\bar{x}z,\bar{x}\bar{z}\}$ and $\{yz,y\bar{z},\bar{y}z,\bar{y}\bar{z}\}$, respectively. The random variable $X$ can be considered in the context $X-Y$ (we denote this random variable by $X_{Y}$). The proposition ``$X=1$ in the context $Y$, disregarding the value of $Y$'', is represented by the proposition $\{xy,x\bar{y}\}$. Its negation, is given by $\{\bar{x}y,\bar{x}\bar{y}\}$. It is easy to check that the set $\mathcal{B}_{X_{Y}}:=\{\emptyset,\{xy,x\bar{y}\},\{\bar{x}y,\bar{x}\bar{y}\},\{xy,x\bar{y},\bar{x}y,\bar{x}\bar{y}\}\}$ forms a Boolean subalgebra of $\mathcal{B}_{X;Y}$. And we also have an isomorphism of Boolean algebras between $\mathcal{B}_{X_{Y}}$ and $\mathcal{P}(\{x,\bar{x}\}):=\mathcal{B}_{X}$. Thus, we have that the random variable $X$ considered in context $Y$ defines a sub-Boolean algebra of $\mathcal{B}_{X;Y}$. The same happens for $Y_{X}$, and $X_{Z}$ with regard to $\mathcal{B}_{X;Z}$, $Y_{Z}$ with regard to $\mathcal{B}_{Y;Z}$, etc. We certainly have that $\mathcal{B}_{X_{Y}}$ is isomorphic to $\mathcal{B}_{X_{Z}}$, $\mathcal{B}_{Y_{X}}$ is isomorphic to $\mathcal{B}_{Y_{Z}}$, etc. Should we identify those random variables? As remarked in the Introduction, this is a crucial problem in probability theory and statistics. In quantum physics, we usually do that, but this is not necessarily so in other fields of research. 

As discussed above, we assume that object's indistinguishability implies the identification of properties. Thus, we assume that $X_{Y}$ and $X_{Z}$ can be identified as random variables. This means that, given the isomorphism between $\mathcal{B}_{X_{Y}}$ and $\mathcal{B}_{X_{Z}}$, for each proposition $F_{1}\in\mathcal{B}_{X_{Z}}$, we have $F_{2}\in\mathcal{B}_{X_{Z}}$ such that its content is the same, and that it has the same probability of occurrence. As an example of this, consider the sets $F_{1}=\{xy,x\bar{y}\}$ (that corresponds to the assertion ``$X=1$ in context $X-Y$'') and $F_{2}=\{xz,x\bar{z}\}$ (that corresponds to the assertion ``$X=1$ in context $X-Z$''). As sets, they are different. But we can identify $F_{1}$ and $F_{2}$ in the following sense: for any (classical) probability assignments $(X-Y,\mathcal{B}_{X;Y}, p_{X;Y})$ and $(X-Z,\mathcal{B}_{X;Z}, p_{X;Z})$, we must have that $ p_{X;Y}(F_{1})=p_{X;Z}(F_{2})$ (i.e., the probabilities are numerically identical for propositions taken from different contexts).

Up to now, we have the following situation. We have three different Boolean algebras of propositions, $\mathcal{B}_{X;Y}$, $\mathcal{B}_{X;Z}$ and $\mathcal{B}_{Y;Z}$.  $\mathcal{B}_{X;Y}$ contains $\mathcal{B}_{X_{Y}}$ and $\mathcal{B}_{Y_{X}}$ as Boolean subalgebras (and the same happens for $\mathcal{B}_{X;Z}$ and $\mathcal{B}_{X_{Z}}$ and $\mathcal{B}_{Z_{X}}$ and $\mathcal{B}_{Y;Z}$ and $\mathcal{B}_{Y_{Z}}$ and $\mathcal{B}_{Z_{Y}}$). Furthermore, we have that, due to the indistinguishability postulate, all probability assignments $(X-Y,\mathcal{B}_{X;Y}, p_{X;Y})$, $(X-Z,\mathcal{B}_{X;Z}, p_{X;Z})$ and $(Y-Z,\mathcal{B}_{Y;Z},p_{Y;Z})$, must be compatible with regard to indistinguishable propositions. Is there a Boolean algebra containing all the propositions in $\mathcal{B}_{X;Y}$, $\mathcal{B}_{X;Z}$ and $\mathcal{B}_{Y;Z}$? Can we find a global probability assignment compatible with $p_{X;Y}$ $p_{X;Z}$ and $p_{Y;Z}$? In the following, we show how to build that required Boolean algebra, and how to build a signed probability assignment for arbitrary (but positive) $p_{X;Y}$ $p_{X;Z}$ and $p_{Y;Z}$.

Define $X-Y-Z:=\{xyz,\bar{x}yz,x\bar{y}z,xy\bar{z},\bar{x}\bar{y}z,\bar{x}y\bar{z},x\bar{y}\bar{z},\bar{x}\bar{y}\bar{z}\}$ and \mbox{$\mathcal{B}_{X;Y;Z}:=\mathcal{P}(X-Y-Z)$}. We need to recover $\mathcal{B}_{X;Y}$, $\mathcal{B}_{X;Z}$ and $\mathcal{B}_{Y;Z}$ as subalgebras of $\mathcal{B}_{X-Y-Z}$. In order to do so, define \mbox{$(X-Y)_{Z}:=\{\{xyz,xy\bar{z}\},\{x\bar{y}z,x\bar{y}\bar{z}\},\{\bar{x}yz,\bar{x}y\bar{z}\},\{\bar{x}\bar{y}z,\bar{x}\bar{y}\bar{z}\}\}$ and $\mathcal{B}_{(X-Y)_{Z}}:=\mathcal{P}((X-Y)_{Z})$}. It~is obvious that $\mathcal{B}_{(X-Y)_{Z}}$ is isomorphic to $\mathcal{B}_{X;Y}$. We can also define $\mathcal{B}_{(X-Z)_{Y}}$ and $\mathcal{B}_{(Y-Z)_{X}}$ in an analogous way, and obtain algebras isomorphic to $\mathcal{B}_{X;Z}$ and $\mathcal{B}_{Y;Z}$, respectively. Similarly, if we consider $\mathcal{B}_{X_{Y-Z}}:=\{\emptyset,\{xyz,x\bar{y}z,xy\bar{z},x\bar{y}\bar{z}\},\{\bar{x}yz,\bar{x}\bar{y}z,\bar{x}y\bar{z},\bar{x}\bar{y}\bar{z}\},\mathbf{1}\}$, we obtain a Boolean subalgebra of $\mathcal{B}_{X;Y;Z}$ which is isomorphic to $\mathcal{B}_{X_{Y}}$. Indeed, $\mathcal{B}_{X_{Y-Z}}$ is isomorphic to $\mathcal{B}_{X_{Y}}$ and $\mathcal{B}_{X_{Z}}$, reflecting the fact that those random variables were identified by the relation ``$\equiv$''.

It is possible now to define a signed probability space $(X-Y-Z,\mathcal{B}_{X;Y;Z},p_{X;Y;Z})$ satisfying Definition \ref{def:negative-probabilities} as follows. Let $p_{X;Y;Z}(F):=p_{X;Y}(F)$, whenever $F\in\mathcal{B}_{(X-Y)_{Z}}$, $p_{X;Y;Z}(F):=p_{X;Z}(F)$, whenever $F\in \mathcal{B}_{(X-Z)_{Y}}$, and $p_{X;Y;Z}(F):=p_{Y;Z}(F)$, whenever $F\in \mathcal{B}_{(Y-Z)_{X}}$. We must also impose that $\sum_{\omega\in X-Y-Z}p_{X;Y;Z}(\omega)=1$. Let us now build $p_{X;Y;Z}$ explicitly. In order to shorten the notation, in~some parts we write $p_{X;Y;Z}(xyz):=p_{xyz}$, $p_{X;Y;Z}(\bar{x}yz):=p_{\bar{x}yz}$, $p_{X;Y;Z}(x\bar{y}z):=p_{x\bar{y}z}$, and so on. The~first constrain that we impose is normalization:
\begin{align}\label{e:NormaSigned}
p_{xyz}+p_{\bar{x}yz}+p_{x\bar{y}z}+p_{xy\bar{z}}+p_{x\bar{y}\bar{z}}+ p_{\bar{x}y\bar{z}}+p_{\bar{x}\bar{y}z}+p_{\bar{x}\bar{y}\bar{z}}=1
\end{align}
Notice that Equation (\ref{e:NormaSigned}) imposes the following normalization conditions on $p_{X;Y}$, $p_{X;Z}$ and $p_{Y;Z}$:
\begin{subequations}
\begin{align}\label{e:normX-Y}
p_{X;Y}(xy)+p_{X;Y}(\bar{x}y)+p_{X;Y}(x\bar{y})+p_{X;Y}(\bar{x}\bar{y})=1
\end{align}
\begin{align}\label{e:normX-Z}
p_{X;Y}(xz)+p_{X;Y}(\bar{x}z)+p_{X;Y}(x\bar{z})+p_{X;Y}(\bar{x}\bar{z})=1
\end{align}
\begin{align}\label{e:normY-Z}
p_{Y;Z}(yz)+p_{Y;Z}(\bar{y}z)+p_{Y;Z}(y\bar{z})+p_{Y;Z}(\bar{y}\bar{z})=1
\end{align}
\end{subequations}

\noindent The context $X-Y$ imposes the following constrains on $p_{X;Y;Z}$. First, notice that $p_{X;Y}$ is fixed by the following: $\langle X\rangle$, $\langle Y\rangle$ and $\langle XY\rangle$, and the normalization condition (\ref{e:normX-Y}). In therms of $p_{X;Y;Z}$, this can be expressed as:

\begin{subequations}\label{e:ContextX-Y}
\begin{align}\label{e:MeanX}
& p_{X;Y;Z}(xyz)-p_{X;Y;Z}(\bar{x}yz)+p_{X;Y;Z}(x\bar{y}z)+p_{X;Y;Z}(xy\bar{z})+p_{X;Y;Z}(x\bar{y}\bar{z})-\\\nonumber
& p_{X;Y;Z}(\bar{x}y\bar{z})-p_{X;Y;Z}(\bar{x}\bar{y}z)-p_{X;Y;Z}(\bar{x}\bar{y}\bar{z})=\langle X\rangle
\end{align}
\begin{align}\label{e:MeanY}
& p_{X;Y;Z}(xyz)+p_{X;Y;Z}(\bar{x}yz)-p_{X;Y;Z}(x\bar{y}z)+p_{X;Y;Z}(xy\bar{z})-p_{X;Y;Z}(x\bar{y}\bar{z})+\\\nonumber
& p_{X;Y;Z}(\bar{x}y\bar{z})-p_{X;Y;Z}(\bar{x}\bar{y}z)-p_{X;Y;Z}(\bar{x}\bar{y}\bar{z})=\langle Y\rangle
\end{align}
\begin{align}
& p_{X;Y;Z}(xyz)-p_{X;Y;Z}(\bar{x}yz)-p_{X;Y;Z}(x\bar{y}z)+p_{X;Y;Z}(xy\bar{z})-p_{X;Y;Z}(x\bar{y}\bar{z})-\\\nonumber
& p_{X;Y;Z}(\bar{x}y\bar{z})+p_{X;Y;Z}(\bar{x}\bar{y}z)+p_{X;Y;Z}(\bar{x}\bar{y}\bar{z})=\langle XY\rangle
\end{align}
\end{subequations}

\noindent Similarly, for the context $X-Z$, besides Equations (\ref{e:MeanX}) and \eqref{e:normX-Z} for the mean value of $X$, we have:

\begin{subequations}\label{e:ContextX-Z}
\begin{align}\label{e:MeanZ}
& p_{X;Y;Z}(xyz)+p_{X;Y;Z}(\bar{x}yz)+p_{X;Y;Z}(x\bar{y}z)-p_{X;Y;Z}(xy\bar{z})-p_{X;Y;Z}(x\bar{y}\bar{z})-\\\nonumber
& p_{X;Y;Z}(\bar{x}y\bar{z})+p_{X;Y;Z}(\bar{x}\bar{y}z)-p_{X;Y;Z}(\bar{x}\bar{y}\bar{z})=\langle Z\rangle
\end{align}
\begin{align}
& p_{X;Y;Z}(xyz)-p_{X;Y;Z}(\bar{x}yz)+p_{X;Y;Z}(x\bar{y}z)-p_{X;Y;Z}(xy\bar{z})-p_{X;Y;Z}(x\bar{y}\bar{z})+\\\nonumber
& p_{X;Y;Z}(\bar{x}y\bar{z})-p_{X;Y;Z}(\bar{x}\bar{y}z)+p_{X;Y;Z}(\bar{x}\bar{y}\bar{z})=\langle XZ\rangle
\end{align}
\end{subequations}

\noindent Finally, for the context $Y-Z$, besides Equation (\ref{e:normY-Z}) and the mean values of $Y$ and $Z$ (given by (\ref{e:MeanY}) and (\ref{e:MeanZ}), respectively), we have
\begin{subequations}
\label{e:ContextY-Z}
\begin{align}
& p_{X;Y;Z}(xyz)+p_{X;Y;Z}(\bar{x}yz)-p_{X;Y;Z}(x\bar{y}z)-p_{X;Y;Z}(xy\bar{z})+p_{X;Y;Z}(x\bar{y}\bar{z})-\\\nonumber
& p_{X;Y;Z}(\bar{x}y\bar{z})-p_{X;Y;Z}(\bar{x}\bar{y}z)+p_{X;Y;Z}(\bar{x}\bar{y}\bar{z})=\langle YZ\rangle
\end{align}
\end{subequations}
Notice that the mean values of $X$, $Y$ and $Z$ are imposed only once. This is possible only because we have made the identifications $X_{Y}\equiv X_{Z}$, $Z_{Y}\equiv Z_{X}$ and $Y_{X}\equiv Y_{Z}$.  Equations (\ref{e:NormaSigned}), (\ref{e:ContextX-Y}), (\ref{e:ContextX-Z}), and~(\ref{e:ContextY-Z}), constitute a set of seven compatible equations for $p_{X;Y;Z}$. As is well known,eight independent equations are needed to define $p_{X;Y;Z}$. Thus, there are infinitely many solutions that satisfy our indistinguishability conditions for contexts. Each one of these solutions, by construction, satisfy our definition of signed probability given in (\ref{def:negative-probabilities}). There is one parameter free for determining $p_{X;Y;Z}$, namely, the mean value $\langle XYZ \rangle$. In order to study the space of solutions, let us write down the matrix form of the set of Equations (\ref{e:NormaSigned}), (\ref{e:ContextX-Y}), (\ref{e:ContextX-Z}), and (\ref{e:ContextY-Z}):

\[
\begin{bmatrix}
% +++ & -++ & +-+ & ++- & +-- &  & -+- & --+ & --- \\
1 & -1 & 1 & 1 & 1 & -1 & -1 & -1 \\
1 & 1 & -1 & 1 & -1 & 1 & -1 & -1 \\
1 & 1 & 1 & -1 & -1 & -1 & 1 & -1 \\
1 & -1 & -1 & 1 & -1 & -1 & 1 & 1 \\
1 & -1 & 1 & -1 & -1 & 1 & -1 & 1 \\
1 & 1 & -1 & -1 & 1 & -1 & -1 & 1 \\
1 & 1 & 1 & 1 & 1 & 1 & 1 & 1 \\
\end{bmatrix}
\begin{bmatrix}
p_{X;Y;Z}(xyz) \\ p_{X;Y;Z}(\bar{x}yz) \\ p_{X;Y;Z}(x\bar{y}z) \\ p_{X;Y;Z}(xy\bar{z}) \\ p_{X;Y;Z}(x\bar{y}\bar{z}) \\  p_{X;Y;Z}(\bar{x}y\bar{z}) \\ p_{X;Y;Z}(\bar{x}\bar{y}z) \\ p_{X;Y;Z}(\bar{x}\bar{y}\bar{z})  
\end{bmatrix}
=
\begin{bmatrix}
\langle X \rangle \\ \langle Y \rangle \\ \langle Z \rangle  \\ \langle XY \rangle \\ \langle XZ \rangle \\ \langle YZ \rangle \\ 1 \\
\end{bmatrix}
\]

The solutions are given by
\begin{subequations}
\begin{equation}
p_{X;Y;Z}(xyz)= \frac{1}{4} \left(1 + \langle XY \rangle + \langle XZ \rangle +\langle YZ \rangle \right)   - \alpha,   
\end{equation}
\begin{equation}
p_{X;Y;Z}(\bar{x}yz)= \frac{1}{4} \left(\langle Y \rangle + \langle Z \rangle -\langle XY \rangle -\langle YZ \rangle \right)  + \alpha , 
\end{equation}
\begin{equation}
p_{X;Y;Z}(x\bar{y}z)= \frac{1}{4} \left(\langle X \rangle + \langle Z \rangle -\langle XY \rangle - \langle YZ \rangle  \right)  +\alpha,
\end{equation}
\begin{equation}
p_{X;Y;Z}(xy\bar{z})= \frac{1}{4} \left(\langle X \rangle + \langle Y \rangle - \langle XZ \rangle - \langle YZ \rangle \right)  + \alpha ,
\end{equation}
\begin{equation}
p_{X;Y;Z}(x\bar{y}\bar{z})=\frac{1}{4} \left(1 - \langle Y \rangle - \langle Z \rangle  + \langle YZ \rangle \right)  - \alpha ,
\end{equation}
\begin{equation}
p_{X;Y;Z}(\bar{x}y\bar{z})=\frac{1}{4} \left(1 - \langle X \rangle - \langle Z \rangle + \langle XZ \rangle \right)  -\alpha ,
\end{equation}
\begin{equation}\label{e:ExaNega}
p_{X;Y;Z}(\bar{x}\bar{y}z)=\frac{1}{4} \left(1 - \langle X \rangle - \langle Y \rangle  +  \langle XY \rangle \right)  - \alpha ,
\end{equation}
\begin{equation}
p_{X;Y;Z}(\bar{x}\bar{y}\bar{z})=\alpha ,
\end{equation}
\end{subequations}
where $\alpha$ is a free parameter. It is immediate from the above solutions that for some correlations, e.g., $\langle XY \rangle =\langle XZ \rangle =\langle YZ \rangle =-1$ no non-negative solutions exist. 

\end{proof}

We use a similar notation as before (but with four jointly measurable pairs) in the following Proposition.

\begin{Proposition}
\label{prop:four-variables}
For jointly measurable pairs $X-Z$, $X-W$, $Y-Z$ and $Y-W$ of dichotomous random variables, if the indistinguishability relations $X_Z\equiv X_W$, $Y_Z\equiv Y_W$, $Z_X\equiv Z_Y$ and $W_X\equiv W_Y$ are satisfied, there exists a signed probability space (i.e., satisfying Definition \ref{def:SignedProba}), for which each pair is a context (satisfying Definition \ref{def:NEWcontext}).
\end{Proposition}

\begin{proof}
Now, let us work out the example with four dychotomic random variables $X$, $Y$, $Z$ and $W$. This example is relevant in the Alice and Bob scenario. Let us assume that $X-Z$, $X-W$ and $Y-Z$ and $Y-W$ form jointly measurable quantities. Proceeding as before, we impose the indistinguishability conditions $X_{Z}\equiv X_{W}$, $Y_{Z}\equiv Y_{W}$, $Z_{X}\equiv Z_{Y}$ and $W_{X}\equiv W_{Y}$. Again, we~will have the Boolean algebras $\mathcal{B}_{X;Z}$, $\mathcal{B}_{X;W}$, $\mathcal{B}_{X;Z}$, $\mathcal{B}_{Y;Z}$, $\mathcal{B}_{Y;W}$, $\mathcal{B}_{X_Z}$, $\mathcal{B}_{X_W}$, and so on. In order to build a Boolean algebra containing all these algebras as subalgebras, consider $X;Y;Z;W:=\{xyzw,\bar{x}yzw,x\bar{y}zw,xy\bar{z}w,xyz\bar{w},\bar{x}\bar{y}zw,\bar{x}y\bar{z}w,\bar{x}yz\bar{w},x\bar{y}\bar{z}w,x\bar{y}z\bar{w},xy\bar{z}\bar{w},\bar{x}\bar{y}\bar{z}w,x\bar{y}\bar{z}\bar{w},\bar{x}y\bar{z}\bar{w},\bar{x}\bar{y}z\bar{w},\bar{x}\bar{y}\bar{z}\bar{w}\}$ and $\mathcal{B}_{X;Y;Z}:=\mathcal{P}(X;Y;Z;W)$. It is straightforward to check that the algebras associated to all jointly measurable variables are subalgebras of $\mathcal{B}_{X;Y;Z}$. Let us work out an example. In order to get a subalgebra of $\mathcal{B}_{X;Y;Z}$ isomorphic to $\mathcal{B}_{X;Z}$, consider the set:
\begin{eqnarray}
\mathcal{P}\left(\{
\{ 
xyzw,
x\bar{y}zw,
xyz\bar{w}, 
x\bar{y}z\bar{w}
\}, 
\{
\bar{x}yzw, 
\bar{x}\bar{y}zw, \bar{x}yz\bar{w},
\bar{x}\bar{y}z\bar{w}
\},\right.
\\ \nonumber 
\left. \{
xy\bar{z}w,
x\bar{y}\bar{z}w,
xy\bar{z}\bar{w},
x\bar{y}\bar{z}\bar{w}
\},
\{
\bar{x}y\bar{z}w,
\bar{x}\bar{y}\bar{z}w,
\bar{x}y\bar{z}\bar{w},
\bar{x}\bar{y}\bar{z}\bar{w}
\}
\}\right)
\end{eqnarray}
Proceeding similarly, we can show that all the desired algebras can be considered as subalgebras of $\mathcal{B}_{X;Y;Z}$. Now, we assume as before that there exist joint probability spaces $(X;Z,\mathcal{B}_{X;Z},p_{X;Z})$, $(X;W,\mathcal{B}_{X;W},p_{X;W})$, $(Y;Z,\mathcal{B}_{Y;Z},p_{Y;Z})$ and $(Y;W,\mathcal{B}_{Y;W},p_{Y;W})$. As before, $(X;Z,\mathcal{B}_{X;Z},p_{X;Z})$ is solely determined by the normalization condition and the values of $\langle X \rangle$, $\langle Z \rangle$ and $\langle XZ \rangle$ (and similar parameters for the other jointly measurable variables). In order to get a global probability, let us proceed us before, by imposing these conditions on $p_{X;Y;Z;W}$. Given that the equations are cumbersome, we just write the matrix equations, which are:
\begingroup\makeatletter\def\f@size{7}\check@mathfonts
\def\maketag@@@#1{\hbox{\m@th\normalsize \normalfont#1}}%
\begin{multline}
\label{eq:AliceBob}
\left( 
\begin{array}{cccccccccccccccc}
% ++++ & -+++ & +-++ & ++-+ & +++- & --++ & -+-+ & -++- & +--+ & ++-- & +-+- & +--- & -+-- & --+- & ---+ & ---- 
1 & -1 & 1  & 1 & 1 & -1 & -1 & -1 & 1  & 1 & 1  & 1  & -1 & -1 & -1 & -1 \\ 
1 & 1  & -1 & 1 & 1 & -1 & 1  & 1  & -1 & 1 & -1 & -1 & 1  & -1 & -1 & -1 \\
1 & 1  & 1  &-1 & 1 & 1  & -1 & 1  & -1 &-1 & 1 & -1 & -1 & 1  & -1 & -1 \\
1 & 1  & 1  & 1 &-1 & 1  &  1 & -1 &  1 &-1 & -1 & -1 & -1 & -1 &  1 & -1 \\
1 & -1 & 1  &-1 & 1 & -1 &  1 & -1 & -1 & -1 &  1& -1 &  1 & -1 &  1 &  1 \\
1 & -1 & 1  & 1 &-1 & -1 & -1 &  1 &  1 & -1 & -1& -1 &  1 &  1 & -1 &  1 \\
1 & 1  & -1 &-1 & 1 & -1 & -1 &  1 &  1 & -1 & -1&  1 & -1 & -1 &  1 &  1 \\
1 & 1  & -1 & 1 & -1& -1 &  1 & -1 & -1 & -1 &  1&  1 & -1 &  1 & -1 &  1 \\
1 & 1  & 1  & 1 & 1 &  1 &  1 &  1 &  1 &  1 &  1&  1 &  1 &  1 &  1 &  1 
\end{array}
\right) \times 
\begin{bmatrix}
p(xyzw) \\ p(\bar{x}yzw) \\ p(x\bar{y}zw) \\ p(xy\bar{z}w) \\ p(xyz\bar{w}) \\  p(\bar{x}\bar{y}zw) \\ p(\bar{x}y\bar{z}w) \\ p(\bar{x}yz\bar{w})\\ p(x\bar{y}\bar{z}w) \\ p(xy\bar{z}\bar{w}) \\ p(x\bar{y}z\bar{w}) \\ p(x\bar{y}\bar{z}\bar{w}) \\ p(\bar{x}y\bar{z}\bar{w}) \\ p(\bar{x}\bar{y}z\bar{w}) \\ p(\bar{x}\bar{y}\bar{z}w) \\ p(\bar{x}\bar{y}\bar{z}\bar{w}) \\
\end{bmatrix}   
= 
\begin{bmatrix}
\langle X \rangle \\ \langle Y \rangle \\ \langle Z \rangle \\ \langle W\rangle  \\ \langle XZ \rangle \\ \langle XW \rangle \\ \langle YZ \rangle \\ \langle YW \rangle \\ 1 \\
\end{bmatrix}
\end{multline}
\endgroup

Each row above corresponds to a linearly independent equation, and therefore the above equations are compatible. Since there are fewer equations than variables, there are infinitely many solutions satisfying our definitions of negative probability and contexts (with seven arbitrary parameters). An~explicit solution is shown in the Appendix \ref{app}.
\end{proof}
The above procedure can be extended to an arbitrary set of dichotomous random variables. Compatible equations are obtained each time we add equations that respect the indistinguishability condition between different random variables.

\section{Conclusions}\label{sec:conclusions}

In this work, we have put forth the following argument. We started by pointing out a well-known and robust connection between contextual theories (such as quantum mechanics) and signed (or negative) probabilities. To generalize this connection, we presented a definition of signed probabilities that relies solely on the notions of signed measurable space and measurement contexts. As expected from previous results, the signed probabilities defined here satisfy the no-signaling condition. With a formal definition of negative probabilities, we followed previous works' reasoning line on indistinguishability and contextuality. We discussed how the assumption of (ontic) particle indistinguishability leads to the following conclusion. Some of the particle testable propositions can be identified among different contexts. This characteristic, in turn, implies the non-signaling condition. Our findings suggest that, in the quantum domain, there is a robust connection between indistinguishability assumptions and the existence of signed probabilities. To~generalize this connection, we presented a definition of signed probabilities that rely on the notions of signed measurable space and measurement contexts, extending Kolmogorov's approach naturally.

It should be clear why negative probabilities are suitable to describe the states of indistinguishable entities. Negative probabilities are necessary and sufficient for no-signaling, and the identification of testable propositions imply no-signaling. Additionally, indistinguishable particles and propositions may lead to contradictions if we assume that their underlying logic is classic. However, as shown in~\cite{de_barros_indistinguishability_2019,de_barros_contextuality_2017}, such contradictions rely on counterfactual reasoning that assumes the classical theory of identity for particles and properties. Therefore, in this situation, we can interpret negative probabilities as the consequence of imposing on indistinguishable particles a classical way of counting, i.e., a~Boolean algebra. When doing so, we need to allow for negative counts to correct for the over-counting of different but indistinguishable particles. This different accounting for events is, in a certain sense, similar to Abramsky and Brandenburger's operational interpretation of negative probabilities \cite{abramsky_operational_2014}. However, contrary to their interpretation, here we propose that this accounting comes from an error in identifying properties, which is due to a fundamental ontological property of particles: they are indistinguishable. 

Paul Dirac was the first to use negative probabilities in physics. He used them to deal with the problem of infinities in quantum field theory \cite{dirac_bakerian_1942}. Later, Richard Feynman tried to use negative probabilities in quantum mechanics \cite{Feynman-1987}. It is fair to say that, though such influential physicists worked with them, negative probabilities remain outside of mainstream physics. The reason is likely not about a lack of meaning for the concept of negative probabilities, as we saw multiple references proposing different interpretations. Perhaps the main reason is that, albeit interesting and easy to compute, negative probabilities did not produce yet any exciting insights into quantum mechanics. We hope that with a well-defined concept of negative probabilities and a connection to a clear ontology inspired by quantum mechanics, negative probabilities can yield new understanding about the quantum world.

\vspace{6pt} 

%%%%%%%%%%%%%%%%%%%%%%%%%%%%%%%%%%%%%%%%%%
%% optional
%\supplementary{The following are available online at \linksupplementary{s1}, Figure S1: title, Table S1: title, Video S1: title.}

\vspace{6pt}

%%%%%%%%%%%%%%%%%%%%%%%%%%%%%%%%%%%%%%%%%%
\authorcontributions{All authors contributed equally for this paper. All authors have read and agreed to the published version of the manuscript.}

%%%%%%%%%%%%%%%%%%%%%%%%%%%%%%%%%%%%%%%%%%
\funding{F.H. was partially funded by the project  “Per un’estensione semantica della Logica Computazionale Quantistica- Impatto teorico e ricadute implementative” , Regione Autonoma della Sardegna, (RAS: RASSR40341), L.R. 7/2017, annualità 2017- Fondo di Sviluppo e Coesione (FSC) 2014--2020.}

%%%%%%%%%%%%%%%%%%%%%%%%%%%%%%%%%%%%%%%%%%
\acknowledgments{Both authors thank Décio Krause and Pawel Kurzynski for discussions. We also thank the anonymous referees for comments and suggestions.}

%%%%%%%%%%%%%%%%%%%%%%%%%%%%%%%%%%%%%%%%%%
\conflictsofinterest{The authors declare no conflict of interest.}

% The following MDPI journals use author-date citation: Arts, Econometrics, Economies, Genealogy, Humanities, IJFS, JRFM, Laws, Religions, Risks, Social Sciences. For those journals, please follow the formatting guidelines on http://www.mdpi.com/authors/references
% To cite two works by the same author: \citeauthor{ref-journal-1a} (\citeyear{ref-journal-1a}, \citeyear{ref-journal-1b}). This produces: Whittaker (1967, 1975)
% To cite two works by the same author with specific pages: \citeauthor{ref-journal-3a} (\citeyear{ref-journal-3a}, p. 328; \citeyear{ref-journal-3b}, p.475). This produces: Wong (1999, p. 328; 2000, p. 475)

%=====================================
% References, variant B: external bibliography
%=====================================
\appendixtitles{no}
\appendix
\section{}\label{app}

Here we write down an explicit solution for the Alice-Bob system of Equations (\ref{eq:AliceBob}). Since (\ref{eq:AliceBob}) has 16 variables but nine equations, the solution will have seven arbitrary parameters, $\alpha_i$, $i=1,\ldots,7$. It is straightforward to compute that a general solution for (\ref{eq:AliceBob}) is the following:
\begin{subequations}

\begin{equation}
p(xyzw)= \alpha_1,
\end{equation}

\begin{equation}
p(xyz\bar{w})= \alpha_2,
\end{equation}
\begin{equation}
p(xy\bar{z}w)= \alpha_3,
\end{equation}
\begin{equation}
p(xy\bar{z}\bar{w})= \alpha_4,
\end{equation}
\begin{equation}
p(x\bar{y}zw)= \alpha_5,
\end{equation}
\begin{equation}
p(x\bar{y}z\bar{w})= \frac{1}{4} \left( 1+ \langle XZ \rangle \right) -\alpha_1 -\alpha_2 -\alpha_5,
\end{equation}
\begin{equation}
p(x\bar{y}\bar{z}w)= \frac{1}{4} \left( 1+ \langle XW \rangle \right) -\alpha_1 -\alpha_3 -\alpha_5,
\end{equation}
\begin{equation}
p(x\bar{y}\bar{z}\bar{w})= -\frac{1}{4} \left(\langle XZ \rangle +\langle XW \rangle \right) +\alpha_1 -\alpha_4 +\alpha_5,
\end{equation}
\begin{equation}
p(\bar{x}yzw)= \frac{1}{4} \left(1 + \langle YW \rangle \right) -\alpha_1 -\alpha_3 +\alpha_6,
\end{equation}
\begin{equation}
p(\bar{x}yz\bar{w})= \frac{1}{4} \left(\langle YZ \rangle -\langle YW \rangle \right) -\alpha_2 +\alpha_6 +\alpha_3,
\end{equation}
\begin{equation}
p(\bar{x}y\bar{z}w)= \alpha_6,
\end{equation}

\begin{equation}
p(\bar{x}y\bar{z}\bar{w})= \frac{1}{4}\left(1 - \langle YZ \rangle  \right) - \alpha_3 - \alpha_4 - \alpha_6,\end{equation}

\begin{equation}
p(\bar{x}\bar{y}zw)= -\frac{1}{4}\left( \langle YW \rangle +\langle XW \rangle \right) +\alpha_1 + \alpha_3 -\alpha_{7} ,
\end{equation}

\begin{equation}
p(\bar{x}\bar{y}z\bar{w})= \frac{1}{4}\left( -\langle XZ \rangle +\langle XW \rangle -\langle YZ \rangle +\langle YW \rangle \right) +\alpha_2 - \alpha_3 +\alpha_7 ,
\end{equation}

\begin{equation}
p(\bar{x}\bar{y}\bar{z}w)= \alpha_7,\end{equation}

\begin{equation}
p(\bar{x}\bar{y}\bar{z}\bar{w})= \frac{1}{4}\left( \langle XZ \rangle + \langle YZ \rangle \right) +\alpha_3 + \alpha_4 - \alpha_{7}.
\end{equation}
It is straightforward to see that for correlations violating the CHSH form of Bell's inequalities, the above solutions cannot be in the interval $[0,1]$, and are therefore not standard probabilities. For example, for the PR-box correlation of $-\langle XZ \rangle= \langle XW \rangle= \langle YZ \rangle= \langle YW \rangle= -1$, it follows that \mbox{$p(x\bar{y}\bar{z}w)=-(\alpha_1+\alpha_3+\alpha_5)$}, which implies that $\alpha_1=\alpha_3=\alpha_5=0$ for it to be non-negative. This~implies, similarly, that $\alpha_7=0$ from $p(\bar{x}yzw)$, $\alpha_2$ from $p(\bar{x}yz\bar{w})$, $\alpha_4$ from $p(x\bar{y}\bar{z}\bar{w})$, $\alpha_6$ from $p(\bar{x}\bar{y}\bar{z}\bar{w})$, and $\alpha_7$ from $p(\bar{x}yzw)$. But since $\alpha_i$ must be zero for $i=1,\ldots,7$, it follows that $p(\bar{x}\bar{y}z\bar{w})=-1/2$, a negative value. Thus, as expected, the PR box maximally violating the CHSH does not have a non-negative joint probability distribution but has a negative probability. Similar results can be obtained for other PR boxes as well as for the QM correlations for the Alice-Bob experiment. 

\end{subequations}

\reftitle{References}
\externalbibliography{yes}

\end{document}